\documentclass[openacc]{rsproca_new}

\renewcommand{\div}[1]{\nabla \cdot {#1}}

\titlehead{Research}
\jname{rspa}
\Journal{Proc R Soc A\ }

\usepackage{graphicx}
\usepackage{newtxtext}
\usepackage{newtxmath}
\usepackage[numbers]{natbib}
\usepackage{hyperref}
\usepackage{caption}
\usepackage{subcaption}

\usepackage{bbold, epsfig}
\usepackage{latexsym, amsmath, euscript,enumitem}
\usepackage[windows]{umlaute}

\begin{document}

\newtheorem{lemma}{Lemma}
\newcommand{\RomanNumeralCaps}[1]

\newcommand{\ds}{\displaystyle}
\renewcommand{\div}{{\rm div \,}}
\newcommand{\grad}{{\rm grad \,}}
\newcommand{\Gr}{\mbox{gr}\,}
\newcommand{\PX}{2^X \setminus \{\emptyset \}}
\newcommand{\sgn}{\,\mbox{sgn}\,}
\newcommand{\Sgn}{\,\mbox{Sgn}\,}
\newcommand{\varchi}{\raisebox{2.5pt}{$\chi$}}
\newcommand{\esssup}[1]{\mathop{\rm ess\ sup}}
\newcommand{\essinf}[1]{\mathop{\rm ess\ inf}}
\newcommand{\N}{{\rm I\kern - 2.5pt N}}
\newcommand{\Z}{{\rm Z\kern - 5.5pt Z}}
\newcommand{\Q}{{\rm I\kern - 5.25pt Q}}
\newcommand{\C}{{\rm I\kern - 6.25pt C}}
\newcommand{\R}{{\rm I\kern - 2.5pt R}}
\newcommand{\D}{\mbox \DH}
\newcommand{\abf}{\mathbf{a}}
\newcommand{\Abf}{\mathbf{A}}
\newcommand{\bbf}{\mathbf{b}}
\newcommand{\Bbf}{\mathbf{B}}
\newcommand{\dbf}{\mathbf{d}}
\newcommand{\Dbf}{\mathbf{D}}
\newcommand{\ebf}{\mathbf{e}}
\newcommand{\Fbf}{\mathbf{F}}
\newcommand{\gbf}{\mathbf{g}}
\newcommand{\Ibf}{\mathbf{I}}
\newcommand{\jbf}{\mathbf{j}}
\newcommand{\Jbf}{\mathbf{J}}
\newcommand{\nbf}{\mathbf{n}}
\newcommand{\Pbf}{\mathbf{P}}
\newcommand{\rbf}{\mathbf{r}}
\newcommand{\Sbf}{\mathbf{S}}
\newcommand{\ubf}{\mathbf{u}}
\newcommand{\vbf}{\mathbf{v}}
\newcommand{\xbf}{\mathbf{x}}
\newcommand{\Phibf}{\mathbf{\Phi}}
\newcommand{\na}{\nabla}
\newcommand{\pa}{\partial}

\title{The initial acceleration of a buoyant spherical bubble revisited}

\author{D. Bothe$^1$,
  J. Liu$^1$,
  P.-E. Druet$^1$,
  T. Maric$^1$,
  M. Niethammer$^1$,
  G. Brenn$^2$}

\address{$^1$Institute of Mathematical Modelling and Analysis, Darmstadt University of Technology, Alarich-Weiss-Stra\ss e 10, 64287 Darmstadt, Germany
$^2$Institute of Fluid Mechanics and Heat Transfer, Graz University of Technology, 8010 Graz, Austria}

\subject{Initial bubble acceleration, multiphase flow, hydrostatic pressure profile}


\corres{G\"unter Brenn\\
\email{guenter.brenn@tugraz.at}}






\begin{abstract}

\textbf{This arXiv preprint is the accepted verion of the published open-access manuscript, please cite the published version linked below:\newline
\href{https://doi.org/10.1098/rspa.2024.0957}{https://doi.org/10.1098/rspa.2024.0957}}

An analytical derivation of the buoyancy-induced initial acceleration of a spherical gas bubble in a host liquid is presented. The theory makes no assumptions further than applying the two-phase incompressible Navier-Stokes equations, showing that neither the classical approach using potential theory nor other simplifying assumptions are needed. The result for the initial bubble acceleration as a function of the gas and liquid densities, classically built on potential theory, is retained. The result is reproduced by detailed numerical simulations. The accelerated, although stagnant state of the bubble induces a pressure distribution on the bubble surface which is different from the result related to the Archimedean principle, emphasizing the importance of the non-equilibrium state for the force acting on the bubble.
\end{abstract}



\begin{fmtext}
\section{Introduction}
\label{sec:intro}
Since the classical work by Poisson \cite{Poisson-1831}, Green \cite{Green-1833}, Stokes \cite{Stokes-1845} and others on the pendulum, it is well-known that, in order to (transiently) move a solid sphere through an ambient fluid, some portion of the host fluid needs to be accelerated to move around the sphere, and then decelerated again. If the particle motion is to be modelled as for a point mass, this requires the use of an effective particle mass, composed of its own mass and the so-called \emph{added} or \emph{virtual mass}. The classical result for a solid sphere, obtained by potential flow theory, states that the added mass equals the mass of ambient fluid within half of the volume of the spherical particle. The derivation is given in many classical texts, such as the treatise by Basset \cite{Basset-1888}. If the particle is a bubble in a surrounding liquid, the added mass is much larger than the mass of the bubble itself. Therefore, if the system is subject to gravity as the body force with initially stag-
\end{fmtext}
\maketitle
\noindent nant fluids, Newton's second law gives approximately, neglecting the bubble's mass, $2g$ as the initial acceleration of the bubble against the direction of gravity.
To substantiate this statement, one needs to extend the derivation such that it applies to a fluid particle, having a deformable and mobile surface, also equipped with capillary force. The passage from solid spheres to spherical fluid particles involves the following steps:
\begin{enumerate}
    \item[1.]~ Account for the particle as a fluid phase composed of a viscous fluid, i.e.\ use appropriate interface conditions for the tangential stress;
    \item[2.] ~Account for the deformability of the fluid particle, i.e.\ treat the interface as an unknown surface that can evolve in the course of time in reaction to the normal stresses acting at the interface;
    \item[3.]~ Account for the convective term, i.e.\ solve the two-phase Navier-Stokes rather than the Stokes equations.
\end{enumerate}

If the convective term in the momentum balance is ignored, a meaningful intermediate step is to rigorously solve the two-phase Stokes equations, i.e.\ to avoid use of potential theory with its underlying assumption of zero vorticity, or of the Stokesian stream function with the underlying assumption of axisymmetry and zero swirl. Alternatively, one can use such specialised approaches if it is shown \emph{a posteriori} that the solution obtained this way actually solves the two-phase Stokes system. For this purpose, the velocity field needs to be complemented by the two-sided pressure field obtained either by solving the Bernoulli equation or by exploiting the pressure gradient obtained from the momentum balance.

Let us note that the incompressible Navier-Stokes equations are appropriate as governing equations as we do not consider phase change. Indeed, the assumption of solenoidal velocity fields, usually called 'incompressibility assumption' in this context, is a very accurate approximation even inside the 
gas phase. 
This is justified by the small Mach number (Ma), starting from ${\rm Ma} =0$ at the initial time $t = 0$.
Hence the velocities are negligible small compared to the speed of sound such that compressibility effects are not present, i.e.\ the flow is isochoric and, hence, the velocity field has vanishing divergence.

Our original motivation to study the initial acceleration of a spherical bubble from rest comes from the need to verify numerical methods and algorithms for two-phase flow simulations. For this purpose, it is very helpful to have appropriate two-phase problems with curved surfaces, for which exact solutions are available. In particular, one is also interested in the local fields of pressure and velocity, in addition to relevant integral quantities. A particular challenge is the proper solution of the two-phase pressure Poisson equation, since the pressure enters the jump relation between normal stress and surface tension forces at the interface. 
For this purpose, we revisited theoretical studies from the literature, as discussed in the survey of section~\ref{sec:lit-survey}, where an acceleration above $2g$ was not to be expected though Dominik \& Cassel reported a value of $3.3g$ \cite{Dominik-Cassel-2015}. The purpose of the present paper is to rigorously derive the initial acceleration of a spherical gas bubble in an infinite ambient liquid from rest \emph{without any further assumptions}, only based on the two-phase incompressible Navier-Stokes equations with the proper interface conditions, positive and constant surface tension, and without phase change. The present paper provides this derivation of the initial acceleration of an individual spherical bubble in an unbounded ambient liquid at rest, together with the associated pressure fields inside and outside the gas-liquid interface. 

In the following section~\ref{sec:lit-survey}, we provide a survey of the published theoretical, experimental and numerical literature on the subject. While several of these publications also consider the bubble rise for a full time interval, the focus of this brief literature review is on the initial acceleration.
Section~\ref{sec:theory} derives an analytical description of the initial bubble acceleration, exploiting the fact that the latter can be computed solely from the pressure field, given that the initial velocity is zero.
We derive the two-phase Poisson equation with jump conditions that applies in this initial time instant and compute the pressure fields in the liquid hosting the accelerated bubble and in the bubble itself. 
Section \ref{sec:simulations} presents results from numerical simulations of a  bubble accelerating in a Newtonian liquid from rest. These results are intended for comparison with the analytical result. 
Finally,  the results are discussed in Section \ref{sec:discussconclus}, where also some conclusions are drawn. 
%
\section{Literature Survey}
\label{sec:lit-survey}
We briefly review the main published contributions concerning the passage from solid to fluid spheres (cf. also the bibliographical note by Pozrikidis \cite{Pozrikidis-1994}). Let us note in passing that in the publications referenced below, the transient motion of the bubble is considered on some time interval of positive length, hence at least for all small times. In contrast to this, the current paper only addresses the initial bubble acceleration, a considerably simpler problem. For the more general topic of forces acting on solid or fluid particles, see the monographs \cite{HappelBrenner1965, Batchelor-1967, Clift-et-al-1978, Michaelides-et-al-2011} and the review articles \cite{Michaelides-2003,Mathai-et-al-2020}. We structure our survey by reviewing theoretical, experimental and numerical work.
\subsection{Theoretical work}\label{subsec:theory-review}
Torobin \& Gauvin \cite{TorobinGauvin1959} presented a series of six papers on fundamental aspects of solid-gas flow. In paper III, the accelerated motion of a particle in a fluid is discussed. The authors conclude that, for describing the accelerated motion, potential flow theory can be used only at the beginning of rectilinear accelerations as well as oscillatory motions involving very small amplitudes. They state that the added-mass concept loses theoretical significance and practical utility if applied beyond this region. Walters \& Davidson \cite{WaltersDavidson1962} considered the initial motion of a gas bubble in 2D, followed by their work \cite{WaltersDavidson1963} in 1963 for the more interesting 3D case. They assumed irrotational flow of the ambient liquid, which was assumed to be an ideal fluid. The problem was treated by means of potential flow theory with assumed axial symmetry. Consequently, zero tangential stress was implicitly imposed as a boundary condition, since an ideal fluid always displays perfect slip at the boundary.  In contrast to a solid sphere, the bubble was assumed to be deformable, where small deformations from an initial spherical shape were treated. But no surface tension was included, so that the force counteracting deformation was not accounted for. Furthermore, by assuming a spatially homogeneous pressure field, the bubble is modelled as a void inside the liquid. Note that this refers to the limit of zero mass density inside the sphere. Expressing the velocity potential as a series of Legendre polynomials with time-dependent coefficients, certain approximations were done to obtain some explicit formulas for the first two coefficient. In particular, under these assumptions they obtained the initial bubble acceleration as $2g$.

Sy et al. considered the transient motion of a gas bubble and a spherical particle under the assumption of creeping flow \cite{Sy-et-al-1970}. They used the Stokes stream function approach, assuming axial symmetry and, thus, implicitly imposing vanishing swirl. Laplace transform techniques are employed to compute and compare the solutions for a solid sphere and an inviscid gas bubble. Under these assumptions, they also considered slight perturbations of the spherical shape, following the approach of Taylor \& Acrivos \cite{Taylor-Acrivos-1964}. They concluded that an initially spherical (gas) bubble with zero density and viscosity will remain spherical in creeping flow, i.e.\ if the convective term can safely be neglected, for all values of the Weber number defined as
\[
We = \frac{R^2 g (\rho^l - \rho^g)}{\sigma (\rho^g / \rho^l + 1/2)},
\]
where $\sigma >0$ was implicitly used. Due to this persistence of the spherical shape, they used the Laplace-transformed solution for the fixed spherical shape, which was approximated and then inverted for small (and, separately, for large) times. In particular, the initial acceleration was calculated to be
\begin{equation}
\abf_B^0 = \frac{\rho^l - \rho^g}{\rho^g + \rho^l /2} \mathbf{g},
\label{eq:ini-acc}
\end{equation}
for both the solid particle and the gas bubble. Since the convective term vanishes at $t = 0$ for a bubble starting from rest, they also concluded that the same result must hold for the initial acceleration of a spherical particle if the ambient flow is described by the Navier-Stokes equations. While this is not a mathematical proof, their arguments are nevertheless plausible.

In a subsequent paper, Sy \& Lightfoot extended the study \cite{Sy-et-al-1970} to account for the viscosity of the gas in the bubble \cite{Sy-Lightfoot-1971}. A spherical shape of constant radius was assumed, such that the normal stress balance at the interface is irrelevant. Again using Stokes stream function calculations, followed by Laplace transform in time, the accelerated motion of the bubble was obtained by numerical inversion of the Laplace transform. Moreover, by an appropriate series representation of the transformed solution, an approximation for short times was obtained, being exact in the limit $t\to 0+$. This way, the initial acceleration was extracted and gave the same result (\ref{eq:ini-acc}). For determining the force on an accelerating body in creeping flow, Morrison \cite{Morrison1972} applied the analysis by Happel \& Brenner \cite{HappelBrenner1965}, who showed that the force can be expressed by means of the stream function of the flow around the body. Using the Laplace transform to solve the unsteady equations of motion, \cite{MorrisonStewart1976} derived the stream function of the starting flow around a small bubble in an accelerating liquid. The authors showed that the governing relation differs from the Basset-Boussinesq-Oseen equation for a solid particle by an integral term accounting for the difference in the interfacial behaviours. Later on, Stewart \& Morrison \cite{StewartMorrison1981} pointed out that the same kinematic viscosity for both phases has erroneously been used in Sy \& Lightfoot \cite{Sy-Lightfoot-1971}. Chisnell also mentions this error \cite{Chisnell-1987}. Moreover, besides giving corrected results and avoiding assumptions from \cite{Sy-et-al-1970} on the density ratio, it is also shown by Chisnell \cite{Chisnell-1987} that the normal stress is homogeneous along the interface in creeping flow. Thus, the bubble stays spherical, indicating that the obtained solution should indeed be a solution of the two-phase Stokes equations. Note that the pressure inside the bubble, which is not considered in \cite{Chisnell-1987}, needs to be adjusted by adding the Laplace pressure in case of positive surface tension. Gordillo et al. \cite{Gordillo-et-al-2012} developed a reduced model for describing the initial state of motion of individual bubbles in stagnant, low-viscosity liquids. The emerging equation for the bubble acceleration as a function of time predicts an initial value of $2g$. The fact that this result does not account for the gas and liquid densities is due to the simplifications underlying the related theory. 

A few years after \cite{Chisnell-1987}, the rigorous mathematical treatment of the incompressible two-phase Navier-Stokes equations with surface tension started. We only mention some most relevant results applicable to a (bounded) capillary fluid particle. The first line of results employs Lagrangian coordinates to reformulate the two-phase Navier-Stokes equations such that the unknown moving interface is replaced by the fixed initial interface. For this to work, the interface has to be a material interface, i.e.\ always formed by the same set of fluid elements. This obviously requires the absence of phase change or, equivalently, that the interface rate of normal displacement equals the normal part of the adjacent bulk phase velocities. This is guaranteed if the interface motion is governed by the kinematic boundary condition, saying that the rate of normal displacement $V_\Sigma$ of the interface equals the adjacent normal bulk velocities. 
Based on results by Solonnikov on the single-phase Navier-Stokes equations and a series of papers on free surface flows \citep{Solonnikov-1986,Solonnikov-1987,Solonnikov-1990,Solonnikov-1991}, Denisova \& Solonnikov obtained the solvability of the incompressible two-phase Navier-Stokes equations in H\"older spaces in a series of papers, where \cite{Denisova-Solonnikov-1995} covers the case of a droplet in infinite three-dimensional space. The proof of existence of such a solution employs successive approximations, using the linearised equations. It is to be noted that the linearisation of the two-phase Navier-Stokes equations can only be done after the interface is fixed, here by passage to Lagrangian coordinates. In the linearisation, the transmission conditions at the interface have to be linearised as well. This leads to additional terms, which are not present in single-phase Navier-Stokes linearisation. These results give rise to classical solutions, which have continuous partial derivatives to the requested order such that all quantities appearing in the partial differential equations and on the boundaries are well-defined in every $(t,x)$. The partial derivatives are actually in certain function spaces of H\"older-continuous functions (H\"older spaces), i.e.\ are somewhat more regular than being just continuous functions. Due to the pointwise fulfilment of all equations, such a solution can only exist if the data of the problem, composed of boundary, transmission and initial conditions (including the initial interface shape), satisfy so-called compatibility conditions. As a simple example note that, if the solution is to satisfy the no slip condition at the domain boundary, then the initial velocity must satisfy the no-slip condition itself. For two-phase incompressible Navier-Stokes equations, these compatibility conditions are somewhat more complicated. Interestingly, it turns out that zero initial velocity fields inside both bulk phases solely fit to a spherical initial interface for a bounded fluid particle, i.e.\ for a bubble or a droplet. This is true since deviations from spherical shape induce surface tension forces that act to bring the shape back to spherical. As these forces are not building up slowly but are instantaneously active, the initial velocity field and the initial interface shape need to be compatible. For full details and a survey on related literature for the Lagrangian approach to the two-phase incompressible Navier-Stokes equations see \cite{Denisova-Solonnikov-2020}.

A different approach treats the system of partial differential equations in the Eulerian formulation. Again, the interface needs to be fixed somehow, introducing further nonlinearities that have to be dealt with to obtain a linearised version. Combining deep knowledge on this linear PDE system with a careful analysis of the nonlinear terms, the solution of the nonlinear complete system is obtained using Banach's fixed point theorem or the implicit function theorem. Fixing the interface in the Eulerian formulation uses the so-called direct mapping method by means of the Hanzawa transform; see \cite{Pruess-Simonett-2016} for full detail. This approach has been employed in \cite{Kohne-Pruess-Wilke-2013} to obtain the existence of a unique solution to the incompressible two-phase Navier-Stokes equations with surface tension, but without body forces. Solvability has been proven in so-called Sobolev-Slobodetskii spaces, where the underlying basic space, in which the partial derivatives typically lie, is the space of $p$-integrable functions for some large $p$. In this case, the solutions satisfy the equations at almost all points in the sense of Lebesgue measure, and no or less compatibility conditions are required, depending on the choice of $p$. Let us note that solutions in this functional analytical setting have also been obtained for the Lagrangian formulation; see \cite{Denisova-Solonnikov-2020}.
The interface can actually be parameterised over a reference surface, the latter being arbitrary close to the initial interface and maximal smooth (a so-called real analytic manifold). This approach allows to prove a regularisation effect due to surface tension: the interface immediately becomes smooth, i.e.\ it is a real analytic surface for every $t>0$; see \cite{Pruess-Simonett-2016}. This approach does not require the interface to be material, and the case of a droplet with evaporation has been studied recently, e.g., in \cite{Pruess-Shimizu-2017}. 

Having these mathematical theorems available allows us to draw some rigorous conclusions concerning the known results coming from simplifying assumptions. For example, the solution of the two-phase Stokes problem for spherical initial shape and zero initial velocity, obtained using Stokes stream function approach, is then the unique solution of this problem. Furthermore, the local in time existence results also make sure that the initial acceleration of the bubble, a quantity we are interested in in the present paper, is well defined for every initial configuration that satisfies certain regularity assumptions and the above-mentioned compatibility conditions. 
In particular, the initial acceleration of a spherical bubble, starting to accelerate in an infinite stagnant liquid, is a well-defined quantity, uniquely determined by the underlying mathematical model.

\subsection{Experimental work}
In the paper with their theoretical analysis, Walters \& Davidson \cite{WaltersDavidson1963} also presented experiments on the rise of bubbles with volumes $110 ml \leq V_b \leq 3000 ml$. They found an initial acceleration of the bubbles of approximately $2g$. Jameson \& Kupferberg studied the formation of individual air bubbles from an orifice covered by a layer of a low-viscosity liquid, treating the bubbles as cylindrical or as spherical \cite{JamesonKupferberg1967}. Motivated by the application to leakage of liquid from sieve trays, the focus of the studies lies on the pressure behind the bubbles after break-off from the source. The pressure is determined as a solution of Bernoulli's equation. The initial bubble acceleration of $2g$ is obtained as a result from this study. Bourrier et al. studied the buoyancy-driven rise of a solid sphere in water, where the density of the sphere (typically a ping-pong ball) was less than that of the water \cite{Bourrier-et-al-1984}. The initial acceleration found from a visualisation of the sphere trajectory equals closely the theoretical result (\ref{eq:ini-acc}), however with the factor $1/2$ in the denominator replaced by an added-mass coefficient $O(0.6)$. Zawala \& Malysa studied the motion of individual gas bubbles coalescing with a free water surface \cite{zawala-malysa-2011}. The individual bubbles were produced by detachment from a submerged capillary orifice. The rising motion of bubbles with four different sizes $O(0.6 mm)$ was studied by visualisation, and the evolution of their positions in time was deduced from the images. The representation of the distance travelled by the bubbles as a second-order polynomial in time until 25 ms after start implied a constant acceleration. The ("initial") acceleration reported is $O(1 g)$, in deviation from many other both theoretical and experimental results. This result shows the difficulty of initial bubble acceleration measurements, thus indicating the importance of numerical simulations. Manica et al. reported theoretical and experimental results about the rise of individual bubbles with immobile or mobile interfaces \cite{Manica-et-al-2016}. In the theoretical analysis, the neglect of the gas density in the bubble against the liquid density leads to the initial bubble acceleration of $2g$, independent of the gas density.

\subsection{Numerical work}
Numerical simulations of the motion and distortion of an individual gas bubble rising through a liquid were presented in \cite{Nilmani-et-al-1981}. The simulated bubbles exhibited volumes between 180 $ml$ and 523 $ml$. Results for the displacement of the bubble centroid as a function of time show that the bubbles initially rise at an acceleration of $2g$. The first author of that paper later investigated the pressure in the liquid behind a rising bubble accelerating from rest \citep{Nilmani-1982}. He showed a zone behind the bubble, where the pressure is reduced as compared to the ambient state. 

In \cite{simcik2008computing}, the authors investigated the possibility of using numerical simulations, performed with the VOF method using the commercial code Fluent, to evaluate the added
mass coefficient of dispersed fluid particles.
For this purpose, the initial acceleration of a buoyant particle,
released from rest, was computed for short initial time intervals, typically
for about $10^{-5}$ s. Different initial shapes were considered. For an initially spherical bubble, an added mass coefficient of 0.5 was found for sufficiently large computational domains to exclude wall effects. This corresponding to an initial acceleration of $1.993g$ for the employed densities of $\rho^l = 998.2$ kg$\, m^{-3}$ and $\rho^g = 1.225$ kg$\, m^{-3}$.

In \cite{Dominik-Cassel-2015}, an inhouse axisymmetric level-set method for incompressible multiphase flows is employed to investigate the initial transient rise of initially spherical gas bubbles in a stagnant liquid.
The numerical method for solving the incompressible two-phase Navier-Stokes equations uses the vorticity-streamfunction formulation.
The authors first recall the classical derivation of the initial  acceleration of a spherical solid, using potential theory and the added-mass concept. Their result for the initial bubble acceleration is $2g$, as the gas density is ignored in their derivation.
This result is compared to the outcome of the level-set simulations, which yield a value of $3.3g$ for the initial acceleration of a spherical air bubble in water, starting from rest.
For a possible explanation, the assumptions underlying the classical derivation are recalled and the classical result is put under doubt. 
Let us note that the numerical method in \cite{Dominik-Cassel-2015}
exploits the assumed axisymmetry of the solution and uses a stream function Ansatz in cylindrical coordinates $(r,x_3)$ to approximate the solution.
To the best of our knowledge, an initial bubble acceleration well above $2g$ was never confirmed by other authors. This actually triggered our search for a rigorous derivation of the initial acceleration directly from the incompressible two-phase Navier-Stokes equations with proper jump conditions, avoiding any simplifying assumption.
%
%
\section{Theory}
\label{sec:theory}
\begin{figure}      
  \centerline{\includegraphics[width=0.5\textwidth]{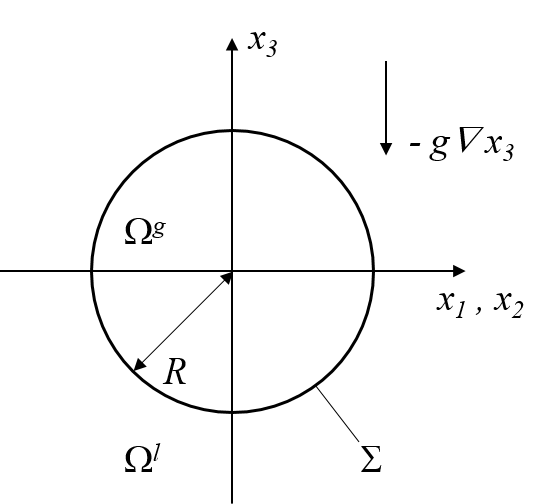}}
  \caption{Individual bubble of radius $R$ in an unbounded liquid host medium. Gravity points in the negative $x_3$ direction.}
\label{fig:sketch}
\end{figure}

The present theory builds on the continuum physical balance equations for mass and momentum inside the bulk phases $\Omega^{g/l}(t)$, with phase indices $g$ and $l$ for gas and liquid, respectively, as well as across the interface $\Sigma (t)$ between them. We assume that the rise of a bubble can be described by the two-phase Navier-Stokes equations for incompressible fluids. The latter assumption means constant mass densities, hence solenoidal flow fields inside the individual bulk phases. Constant mass density is an excellent approximation for the liquid phase under the assumed isothermal conditions, while incompressibility of the gas phase is
based on the low Mach-number approximation, being valid for small velocities as compared
to the speed of sound in the medium. Accordingly, the governing equations inside the bulk phases $\Omega^{g/l}(t)$ read as
\begin{equation}
	\na \cdot \vbf=0,\label{E1}\quad
	\rho \pa_t \vbf+ \rho (\vbf \cdot \na ) \vbf =\na \cdot \Sbf +\rho\bbf
\end{equation}
with the stress tensor
\begin{equation}\label{E3}
\Sbf=-p \Ibf+\Sbf^{\rm visc}
\; \mbox{ with } \Sbf^{\rm visc}=\eta(\na\vbf+\na \vbf^{\sf T})
\end{equation}
and the specific body force $\bbf$. The material parameters $\rho$ (density) and $\eta$ (dynamic viscosity) depend on the respective phase. For instance, $\rho= \rho^g$ in the gas and $\rho= \rho^l$ in the liquid phase.
At the interface $\Sigma (t)$, which separates the two bulk phases $\Omega^g (t)$ and $\Omega^l (t)$,
all the field quantities are assumed to have one-sided limits, which are related via jump conditions (cf. Fig. \ref{fig:sketch}).
The interfacial jump conditions for mass and momentum are
\begin{equation}\label{E4}
[\![\vbf]\!]=0, \quad [\![-\Sbf]\!] \cdot \nbf_\Sigma=\sigma \kappa_\Sigma \nbf_\Sigma,
\end{equation}
where $\nbf_\Sigma$ is the unit normal and $\kappa_\Sigma=- \nabla_\Sigma \cdot \nbf_\Sigma$ is twice the (local) mean curvature of the interface, i.e.\ $\kappa_\Sigma \equiv 2 /R$ for $\Sigma$ given as a sphere of radius $R>0$ with inward orientation of the interface normal $\nbf_\Sigma$ (in order to obtain a positive mean curvature for the sphere).
The jump bracket $[\![\phi]\!]$ denotes the jump of a field $\phi$ across the interface, $[\![\phi]\!] = \phi^g - \phi^l$ in case $\nbf_\Sigma$ is oriented as mentioned above. Note that we assume constant surface tension $\sigma$ throughout; otherwise, the additional Marangoni stress given by $\nabla_\Sigma \sigma$ appears on the right-hand side of \eqref{E4}. Furthermore, no phase change is present, hence the interface is passively advected with the bulk flow fields, i.e.\
\begin{equation}\label{kinematic-BC}
V_\Sigma = \vbf \cdot \nbf_\Sigma,
\end{equation}
where $V_\Sigma$ denotes the speed of normal displacement of the interface.
Equation (\ref{kinematic-BC}) is the kinematic boundary condition. Below, an important consequence of (\ref{kinematic-BC}) is exploited: throughout the motion of the bubble, the interface $\Sigma (t)$ is always composed of the same fluid particles. In other words,
given a point $\xbf_0 \in \Sigma (t_0)$, 
the (unique) solution of the initial value problem 
\begin{equation}\label{kinematic-BC-2}
\dot{\xbf} (t) = \vbf (t, \xbf (t)), \quad  \xbf (t_0) = \xbf_0
\end{equation}
satisfies $\xbf (t) \in \Sigma (t)$ for all $t$. A proof of this fact can be found, e.g., in \cite{bothe2020moving}.

Together with appropriate initial and boundary conditions, the two-phase Navier-Stokes equations describe the hydrodynamics of a rising bubble in a Newtonian liquid without phase change under isothermal conditions. To study the initial acceleration of a spherical bubble under buoyancy from rest, we consider as initial conditions:
\begin{equation}\label{E5}
\Omega^g (0)=B_R(0),\quad \Omega^l(0)=\R^3 \setminus \bar{B}_R(0), \quad \Sigma(0)=\partial B_R (0), \quad
\vbf (0, \cdot)=0.
\end{equation}
As we are interested in the bubble acceleration, we first let
\begin{equation}\label{E6}
\xbf_B (t)=\frac{1}{|\Omega^g|} \int_{\Omega^g(t)} \xbf \,d\xbf
\end{equation}
denote the position of the bubble, where $|\Omega^g|=\frac 4 3 \pi R^3$ is the constant volume of the bubble.
This yields the bubble velocity as
\begin{equation}\label{E6a}
\vbf_B (t)=\dot{\xbf}_B (t)
=\frac{1}{|\Omega^g|} 
\int_{\partial \Omega^g(t)} \xbf \,  (\vbf \cdot \nbf) \,do
=\frac{1}{|\Omega^g|} \int_{\Omega^g(t)} \vbf \,d\xbf,
\end{equation}
where the Reynolds transport theorem has been employed and
$\nbf$ is the outward unit normal to $\Omega^g(t)$.
Another application of the Reynolds transport theorem yields the
bubble's acceleration according to
\begin{equation}\label{E6b}
\abf_B (t)=\dot{\vbf}_B(t)
=\frac{d}{dt} \frac{1}{|\Omega^g|} \int_{\Omega^g(t)} \vbf \,d\xbf
=\frac{1}{|\Omega^g|} \int_{\Omega^g(t)} 
\frac{1}{\rho^g} \big( \na \cdot \Sbf + \rho^g \bbf \big) \,d\xbf.
\end{equation}
The only body force considered is the gravitational force, acting in the negative $x_3$-direction, i.e.\ $\bbf = - g \na x_3$ with $g=9.81 \, {\rm m\, s}^{-2}$. This yields
\begin{equation}\label{E6d}
\abf_B (t)=\frac{1}{|\Omega^g|} 
\int_{\partial \Omega^g(t)} \frac{1}{\rho^g} 
 \Sbf \cdot \nbf \,do - g \ebf_3,
\end{equation}
where $\ebf_3$ denotes the unit vector in the positive $x_3$-direction.
As the initial velocity vanishes, taking the limit as $t\to 0+$ yields
\begin{equation}\label{E6e}
\abf_B^0 :=\abf_B (0+)=- \,\frac{1}{|\Omega^g|} 
\int_{|\xbf |=R} \frac{p^g}{\rho^g} \,  \frac{\xbf}{R} \,do  - g \ebf_3.
\end{equation}
We thus need to compute the initial pressure field.
As we assume purely hydrostatic pressure far away from the bubble, we introduce a modified pressure $\pi$ according to
\begin{equation}\label{E7}
\pi^g = p^g + \rho^g g x_3 - \frac{2 \sigma}{R}, \quad \pi^l = p^l + \rho^l g x_3,
\end{equation}
where we also subtracted the Laplace pressure jump inside the gas bubble to further simplify the momentum jump condition for the initial time instant. In terms of the modified pressure, the latter now reads as
\begin{equation}\label{E8}
[\![\pi]\!]= 2 \langle (\eta^g \na \vbf^g - \eta^l \na \vbf^l) \cdot \nbf_\Sigma, \nbf_\Sigma \rangle + [\![\rho]\!] g x_3 + \sigma (\kappa_\Sigma - \frac 2 R ),
\end{equation}
where the last term vanishes at $t=0$.
The bulk momentum balance becomes
\begin{equation}\label{E9}
\rho \pa_t \vbf + \rho  (\vbf \cdot \na)  \vbf = - \na \pi + \eta \Delta \vbf.
\end{equation}
Taking the divergence of \eqref{E9} and exploiting the fact that $\vbf$ is solenoidal, we see that the modified pressure satisfies the Poisson equation
\begin{equation}\label{E10}
	- \Delta \pi = \rho \na \vbf : \na \vbf.
\end{equation}
Notice that the pressure needs to satisfy a two-phase Poisson problem, hence a further jump condition is required.
As \eqref{E8} is a Dirichlet-type jump condition, we look for an additional Neumann-type transmission condition.
For this purpose, we divide \eqref{E9} by the (phase specific) mass density $\rho$ and
take the inner product with the interface normal. Then, taking the difference of the one-sided limits at the interface, we obtain
\begin{equation}\label{E11a}
[\![ \pa_t \vbf +  (\vbf \cdot \na)  \vbf  ]\!]\cdot \nbf_\Sigma
+ [\![ \frac 1 \rho  \frac{\pa \pi}{\pa \nbf_\Sigma}  ]\!]
=[\![ \frac{\eta}{\rho}  \Delta \vbf ]\!] \cdot \nbf_\Sigma.
\end{equation}
The term inside the first jump bracket is the Lagrangian derivative $\frac{D\vbf}{Dt}$ of the velocity field, and this derivative is well-defined at the interface as the fluid trajectories stay inside $\Sigma (\cdot)$. Moreover,
the continuity of the velocity at $\Sigma$, i.e.\ (\ref{E4})$_1$, implies $\frac{D\vbf^g}{Dt} = \frac{D\vbf^l}{Dt}$ on $\Sigma$. Consequently, (\ref{E11a})
becomes
\begin{equation}\label{E11}
[\![ \frac 1 \rho  \frac{\pa \pi}{\pa \nbf_\Sigma}  ]\!]
=[\![ \frac{\eta}{\rho}  \Delta \vbf ]\!] \cdot \nbf_\Sigma.
\end{equation}

As we are only interested in the initial acceleration from zero velocity, we consider the time instant $t=0$, more precisely the limit as $t\to 0+$. We can then simplify the problem further by exploiting the vanishing initial velocity field.
We write $\pi$ instead of $\pi(0+,\cdot)$.
Then the initial pressure field solves the two-phase Poisson problem
\begin{align}\label{E13}
\Delta \pi & = 0 & \mbox{ for } |x|\neq R,\\[0.5ex]
[\![\pi]\!] & = [\![\rho]\!] g x_3 & \mbox{ for } |x|=R,\label{E13b}\\
[\![ \frac 1 \rho \frac{\pa \pi}{\pa \nbf_\Sigma}  ]\!] & = 0 & \mbox{ for } |x|=R,\label{E13c}\\
\pi & \to 0 & \mbox{ for } |x|\to \infty.\label{E13d}\hspace{-0.1in}
\end{align}
It is not difficult to show that this problem has a unique solution
in the class of functions being twice continuously differentiable with sufficiently fast decay at infinity.
Indeed, if $\pi$ and $\tilde{\pi}$ are two solutions of (\ref{E13}) - (\ref{E13d}),
their difference $u:=\pi - \tilde{\pi}$ satisfies (\ref{E13}) - (\ref{E13d}) with zero right-hand sides. Then, integrating $\frac 1 \rho u\, \Delta u =0$ over $B_r (0)$ for $r>R$ and applying partial integration, we have
\begin{equation}
0 = \int_{B_{r} (0)} \frac u \rho \Delta u \, d\xbf =
\int_{|\xbf|=r} \frac u \rho \frac{\pa u}{\pa \nbf} do
-\int_{B_{r} (0)} \frac 1 \rho | \nabla u |^2 d\xbf
- \int_{|\xbf|=R} [\![ \frac u \rho \frac{\pa u}{\pa \nbf_\Sigma}  ]\!] do,
\end{equation}
where $\nbf$ denotes the outer normal at $|\xbf|=r$.
Exploiting the homogeneous jump condition $[\![ u ]\!]=0$ inherited from (\ref{E13b}), i.e.\ the continuity of $u$ at $\Sigma$, gives
\[
[\![ \frac u \rho \frac{\pa u}{\pa \nbf}  ]\!]=
u\, [\![ \frac 1 \rho \frac{\pa u}{\pa \nbf}  ]\!]=0,
\]
where (\ref{E13c}) for $u$ instead of $\pi$ is used as well.
Taking the limit as $r\to \infty$, we obtain
\begin{equation}
\int_{\R^3} \frac 1 \rho | \nabla u |^2 d\xbf
=\lim_{r\to \infty} \int_{|\xbf|=r} \frac u \rho \frac{\pa u}{\pa \nbf} do =0,
\end{equation}
as $u\to 0$ at infinity due to (\ref{E13d}) and $\nabla u$ is bounded due to the assumption of fast decay of $u$ to zero.
Hence $u\equiv const =0$, where the last equality comes again from (\ref{E13d}).
Thus, $\pi = \tilde{\pi}$, i.e.\ any two solutions coincide.

We are hence done if we just find any classical solution of this two-phase
Poisson problem.
As the pressure jump along the interface depends linearly on $\xbf$, we try the Ansatz
\begin{align}\label{E14}
\pi^g  & = \langle \Abf^g , \xbf \rangle + B^g & \mbox{ for } |x|<R,\\
\pi^l  & = \langle \Abf^l , \xbf \rangle (R/|\xbf |)^3 + B^l R/|\xbf | & \mbox{ for } |x|>R\;
\end{align}
with unknown coefficients $\Abf^g, \Abf^l \in \R^3$ and $B^g, B^l \in \R$.
By straightforward computation, we obtain
\begin{align}\label{E15}
[\![\pi]\!]  & = \langle \Abf^g - \Abf^l , \xbf \rangle  + B^g - B^l & \mbox{ at } |x|=R,\\
[\![ \frac 1 \rho \frac{\pa \pi}{\pa \nbf_\Sigma}  ]\!]  & 
= - \langle \frac{1}{ \rho^g} \Abf^g + \frac{2}{\rho^l} \Abf^l ,\frac \xbf R \rangle - \frac{B^l }{\rho^l R}   & \mbox{ at } |x|=R.
\end{align}
Insertion into the jump conditions shows them to be equivalent to $B^g=B^l =0$ and
$\Abf^g - \Abf^l = [\![\rho]\!] g\, \ebf_3 $ and $\Abf^g /\rho^g+ 2 \Abf^l / \rho^l =0$, hence
\begin{equation}\label{E16}
\Abf^g = [\![\rho]\!] \frac{2 \rho^g}{\rho^l + 2 \rho^g} g\, \ebf_3, 
\quad \Abf^l = - [\![\rho]\!] \frac{\rho^l}{\rho^l + 2 \rho^g} g\, \ebf_3, \quad
B^g=B^l =0.
\end{equation}
Inserting the results from (\ref{E16}) into the Ansatz from above and using
the relation (\ref{E7}) between $\pi$ and $p$ gives the pressure inside the bulk phases as
\begin{align}
p^g (\xbf) & =  -\frac{3 \rho^l}{\rho^l + 2 \rho^g} \rho^g g \, x_3 + \frac{2 \sigma}{R} & \mbox{ for } |x|<R,\label{eq:pgas-fin}
\\ 
p^l (\xbf) & =  -\, \left[ \frac{\rho^g - \rho^l}{\rho^l + 2 \rho^g} \left( \frac{R}{|\xbf|} \right)^3 + 1 \right] \rho^l g \, x_3 & \mbox{ for } |x >R. \label{eq:pliquid-fin}
\end{align}
Note that the pressure $p^g(\xbf)$ in the gas phase varies with the vertical coordinate $x_3$ in a manner different from the equilibrium hydrostatic case. For a liquid density much greater than the gas density, we observe in the gas pressure gradient in the vertical direction a factor of $3$ in this solution as compared to the value of unity to be expected for the equilibrium hydrostatic state. The associated acceleration, together with the gravitational acceleration $g$ in the opposite direction accounting for the bubble weight, represents correctly the resultant bubble acceleration of approximately $2g$. The pressure in the liquid phase varies with the distance from the bubble center and the vertical coordinate $x_3$ in a manner that the equilibrium hydrostatic pressure profile is reached at a large distance from the bubble. For a ratio $\rho^g / \rho^l = 10^{-3}$, it takes a distance from the bubble center of approximately $5~R$ for the pressure profile $p^l(\xbf)$ to deviate from the equilibrium profile by no more than 1\%. The pressure gradient close to the bubble surface, divided by the density, balances the other forces acting, so that in the gas phase, where $1/\rho^g$ has a large value, the corresponding pressure gradient is small. Close to the interface, therefore, the pressure in the liquid phase varies only slightly.
\begin{figure}      
  \centerline{\def\svgwidth{.6\textwidth}
    {\scriptsize
    \graphicspath{{figures/}}
\begingroup%
  \makeatletter%
  \providecommand\color[2][]{%
    \errmessage{(Inkscape) Color is used for the text in Inkscape, but the package 'color.sty' is not loaded}%
    \renewcommand\color[2][]{}%
  }%
  \providecommand\transparent[1]{%
    \errmessage{(Inkscape) Transparency is used (non-zero) for the text in Inkscape, but the package 'transparent.sty' is not loaded}%
    \renewcommand\transparent[1]{}%
  }%
  \providecommand\rotatebox[2]{#2}%
  \newcommand*\fsize{\dimexpr\f@size pt\relax}%
  \newcommand*\lineheight[1]{\fontsize{\fsize}{#1\fsize}\selectfont}%
  \ifx\svgwidth\undefined%
    \setlength{\unitlength}{411.0474115bp}%
    \ifx\svgscale\undefined%
      \relax%
    \else%
      \setlength{\unitlength}{\unitlength * \real{\svgscale}}%
    \fi%
  \else%
    \setlength{\unitlength}{\svgwidth}%
  \fi%
  \global\let\svgwidth\undefined%
  \global\let\svgscale\undefined%
  \makeatother%
  \begin{picture}(1,0.84289371)%
    \lineheight{1}%
    \setlength\tabcolsep{0pt}%
    \put(0,0){\includegraphics[width=\unitlength,page=1]{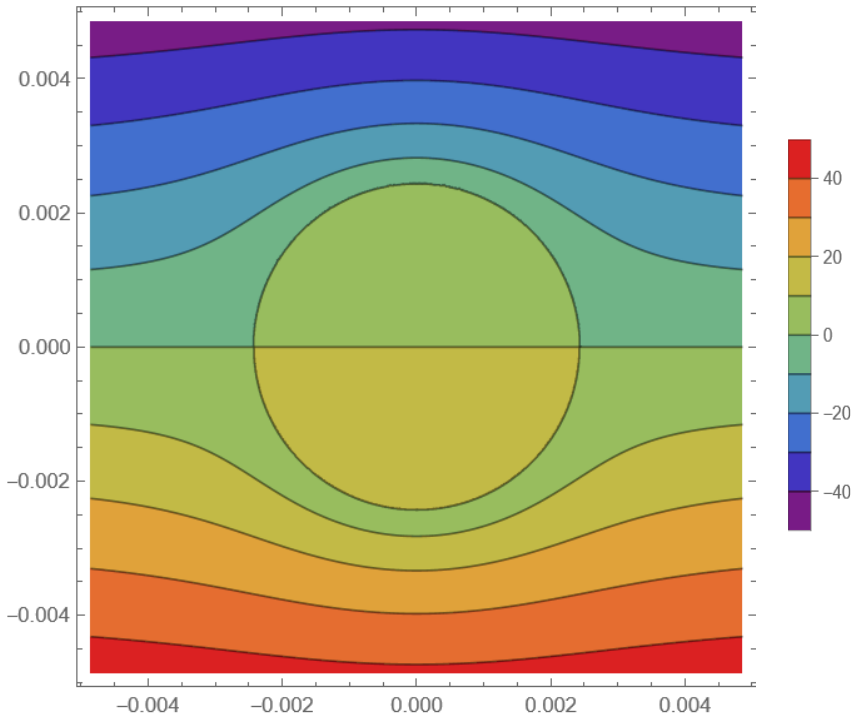}}%
    \put(1.04753429,0.43249939){\color[rgb]{0,0,0}\rotatebox{90}{\makebox(0,0)[t]{\lineheight{13.25}\smash{\begin{tabular}[t]{c}Pressure (Pa) \end{tabular}}}}}%
    \put(0.48382195,-0.01135562){\color[rgb]{0,0,0}\makebox(0,0)[t]{\lineheight{13.25}\smash{\begin{tabular}[t]{c}$x_2$\end{tabular}}}}%
    \put(0.00518834,0.43390165){\color[rgb]{0,0,0}\rotatebox{90}{\makebox(0,0)[t]{\lineheight{13.25}\smash{\begin{tabular}[t]{c}$x_3$\end{tabular}}}}}%
  \end{picture}%
\endgroup%
}
  }
  \caption{Pressure in the gas and liquid phases in the $x_2, x_3$-plane (at $x_1 = 0$) for a bubble accelerated from rest. Analytical pressure field (in $Pa$) at $t=0\,s$. A Laplace pressure jump of $10\, Pa$ is imposed in order to distinguish the bubble contour. Coordinate values in $m$.}
\label{fig:inipress}
\end{figure}
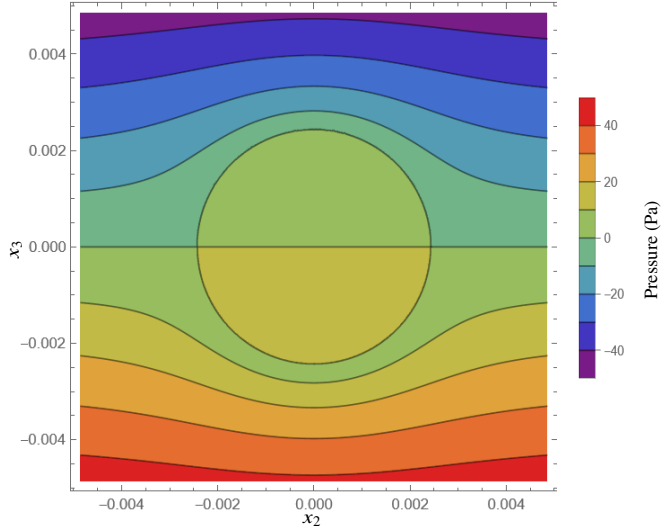
To obtain the initial acceleration, we insert the pressure $p^g$ from
(\ref{eq:pgas-fin}) into (\ref{E6e}) and get
\begin{equation}\label{E17}
\abf_B^0 = \frac{1}{|\Omega^g|} 
\int_{|\xbf |=R} 
\frac{3 \rho^l}{\rho^l + 2 \rho^g} g \, x_3 
\,  \frac{\xbf}{R} \,do  - g \ebf_3.
\end{equation}
This yields
\begin{equation}\label{E22}
\abf_B^0 =
2 \frac{\rho^l - \rho^g}{\rho^l + 2 \rho^g} g\, \ebf_3.
\end{equation}
This is identical to the classical result (\ref{eq:ini-acc}), obtained by potential theory; cf.\ subsection~\ref{subsec:theory-review} above.
The present derivation shows that neither potential theory nor other simplifications are needed for obtaining it. The only underlying assumption is the spherical initial shape of the bubble. A good approximation of the value of the initial bubble acceleration for an air bubble in water is $a_B^0 = 2 g$.

The pressure field from equations (\ref{eq:pgas-fin}) and (\ref{eq:pliquid-fin}) is displayed in Fig. \ref{fig:inipress} in the form of a contour plot, showing the meridional cut through the bubble surface at $x_1 = 0$ by a circle. 
An artificial surface tension value was used such that the Laplace pressure jump amounts to $10\; Pa$. This allows to detect the bubble contour, while avoiding pressure values inside the bubble being out of scale. Recall that the value of the (constant) surface tension is irrelevant for the initial acceleration as it does not induce a resultant force on the bubble. Note that the pressure profiles in both the gas and the liquid phase found for this accelerated state deviate substantially from the equilibrium hydrostatic case.
Let us also note in passing that, while the initial acceleration in (\ref{E22}) can be found in several pieces of literature, the pressure field around the accelerating fluid particle is usually not given explicitely.

\section{Numerical simulations}
\label{sec:simulations}

We numerically solve the incompressible two-phase Navier-Stokes equations (\ref{E1}, \ref{E4}, \ref{kinematic-BC}) with the plicRDF-isoAdvector \cite{Roenby2016,Scheufler2019} geometric volume-of-fluid (VOF) method, while ensuring consistency between volume, mass and momentum conservation \cite{Liu2024rhoVOF} using a consistent choice of discretization schemes. The phase indicator function 
\begin{equation}
  \chi(t, \mathbf{x}) = 
    \begin{cases}
      1 & \text{for } \mathbf{x} \in \Omega^l(t), \\ 
      0 & \text{for } \mathbf{x} \in \Omega^g(t),
    \end{cases}
  \label{eq:indicator}
\end{equation}
is employed to distinguish between the phases. The phase indicator advection is governed by the transport equation
\begin{equation}
    \pa_t \chi  + \vbf \cdot \nabla \chi =0.
    \label{eq:chi}
\end{equation}
Equation (\ref{eq:chi}) must be solved numerically together with equations (\ref{E1}, \ref{E4}, \ref{kinematic-BC}). To solve (\ref{eq:chi}) using the unstructured geometrical VOF method, we define the volume fraction as
\begin{equation}
    \alpha_c(t):=\frac{1}{|\Omega_c|}\int_{\Omega_c} \chi(\mathbf{x}, t) d\xbf
    \label{eq:volfracdef}
\end{equation} 
for an arbitrary fixed control volume $\Omega_c$, and use equation (\ref{eq:volfracdef}) to reformulate equation (\ref{eq:chi}) into an integral conservative transport equation for the volume fraction $\alpha_c$ and the phase indicator $\chi$ within $\Omega_c$, i.e.,
\begin{equation}
    \pa_t \alpha_c = - \frac{1}{|\Omega_c|}\int_{\partial\Omega_c} \chi \vbf\cdot \nbf \, do,
    \label{eq:vofint}
\end{equation}
where $\partial\Omega_c$ is the boundary of $\Omega_c$
and $\nbf$ is the outer normal for $\Omega_c$; see the review article
\citep{Maric2020review} for more details.
The plicRDF-isoAdvector method solves equation (\ref{eq:vofint}) by approximating the phase indicator $\chi$ inside $\Omega_c$ based on piecewise linear interface calculation (PLIC), which  iteratively reconstructs a planar signed distance function associated to centers of finite volumes $\Omega_c$ for which $0 < \alpha_c < 1$, and cells that share at least one point with $\Omega_c$. An iteratively reconstructed signed distance function increases the convergence rate of approximated interface normal vectors. With a more accurately reconstructed interface, the volume fraction advection scheme isoAdvector furthers a high degree of local volume conservation, very low numerical diffusion of volume fractions, and intrinsic handling of topological changes of the fluid interface, making it particularly effective for capturing complex multiphase hydrodynamics. Although the governing equations behind the VOF method consistently transport density, volume and momentum, they must be carefully discretized to maintain this consistency for phases with strongly different densities \citep{Liu2023rhoLENT,Liu2024rhoVOF}. 

\begin{figure}      
\centerline{
\def\svgwidth{.33\textwidth}
    {\scriptsize
    \graphicspath{{figures/}}
     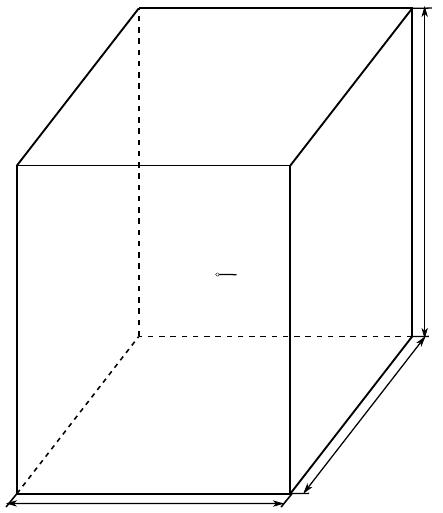
    }  (a) \hspace{1.2em}
\includegraphics[width=0.6\textwidth]{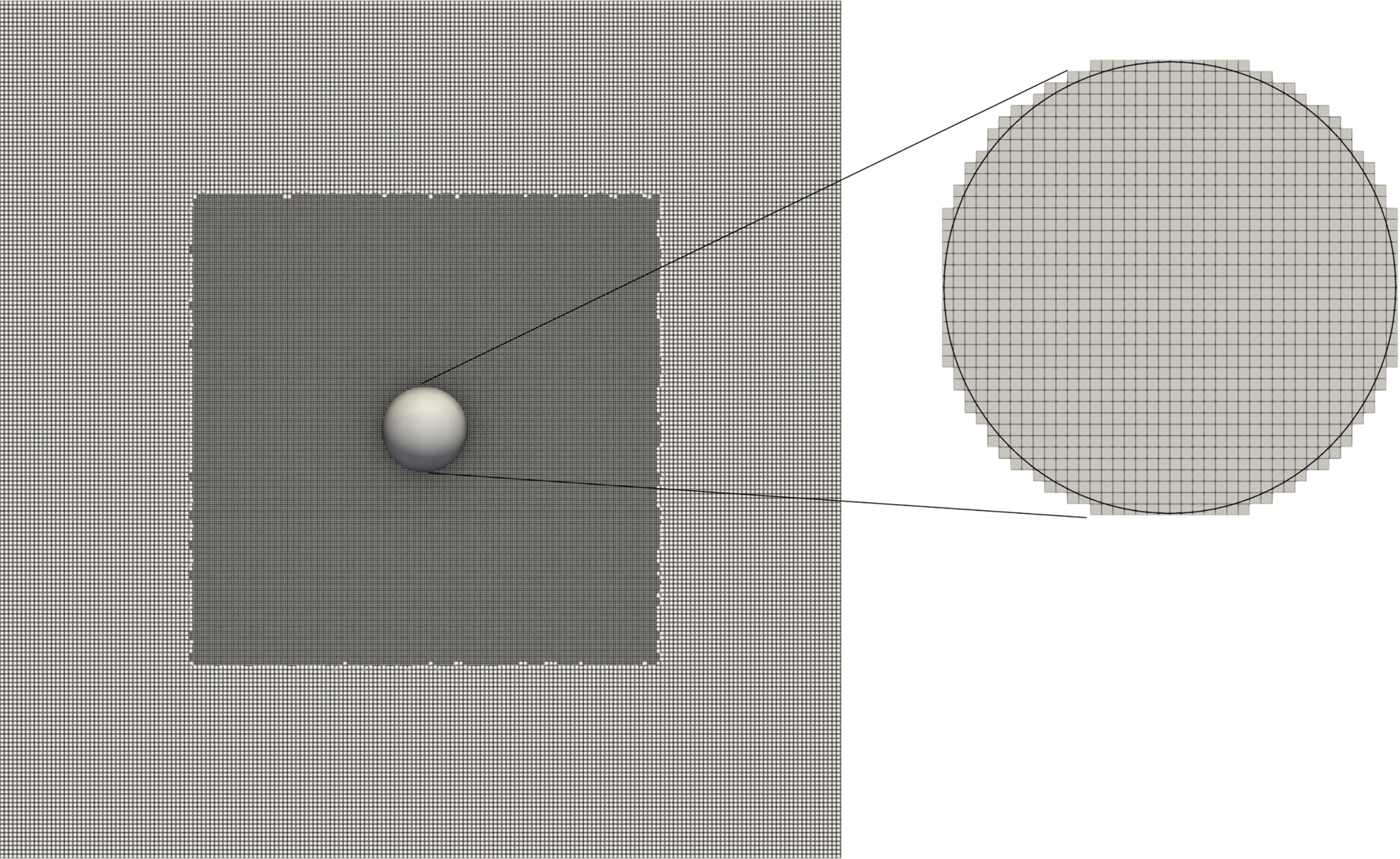} (b)     
    } 
\caption
{Computational domain and grid for the numerical simulations of the initial accelerated state of motion of an individual bubble:
(a) Computational domain given as the cube $(-10R, 10R)^3$ with the center of the bubble with radius $R=2.43mm$ initially located at the origin; gravity points in the negative $x_3$ direction; (b) mesh with one refinement level and a mesh size of $h=\frac{D}{40}$ in the bubble surrounding region.}
\label{fig:sketchnum}
\end{figure}

Numerical simulations of the initial acceleration of a single air bubble in water confirm the analysis of Section~\ref{sec:theory}. Underlying assumptions include that the bubble starts accelerating as a perfect sphere and that the vector of gravitational acceleration points into the negative $x_3$ direction of a Cartesian coordinate system (cf. Figure~\ref{fig:sketchnum}$\,$(a)). The bubble's volume is set to $60\,mm^3$, corresponding to a diameter of $D=4.86\,mm$. The initial center of the bubble was positioned at the origin of the coordinate system.
\begin{figure}      
  \centerline{
      \def\svgwidth{.45\textwidth}
    {\scriptsize
    \graphicspath{{figures/}}
\begingroup%
  \makeatletter%
  \providecommand\color[2][]{%
    \errmessage{(Inkscape) Color is used for the text in Inkscape, but the package 'color.sty' is not loaded}%
    \renewcommand\color[2][]{}%
  }%
  \providecommand\transparent[1]{%
    \errmessage{(Inkscape) Transparency is used (non-zero) for the text in Inkscape, but the package 'transparent.sty' is not loaded}%
    \renewcommand\transparent[1]{}%
  }%
  \providecommand\rotatebox[2]{#2}%
  \newcommand*\fsize{\dimexpr\f@size pt\relax}%
  \newcommand*\lineheight[1]{\fontsize{\fsize}{#1\fsize}\selectfont}%
  \ifx\svgwidth\undefined%
    \setlength{\unitlength}{1157.66927419bp}%
    \ifx\svgscale\undefined%
      \relax%
    \else%
      \setlength{\unitlength}{\unitlength * \real{\svgscale}}%
    \fi%
  \else%
    \setlength{\unitlength}{\svgwidth}%
  \fi%
  \global\let\svgwidth\undefined%
  \global\let\svgscale\undefined%
  \makeatother%
  \begin{picture}(1,0.78232127)%
    \lineheight{1}%
    \setlength\tabcolsep{0pt}%
    \put(0,0){\includegraphics[width=\unitlength,page=1]{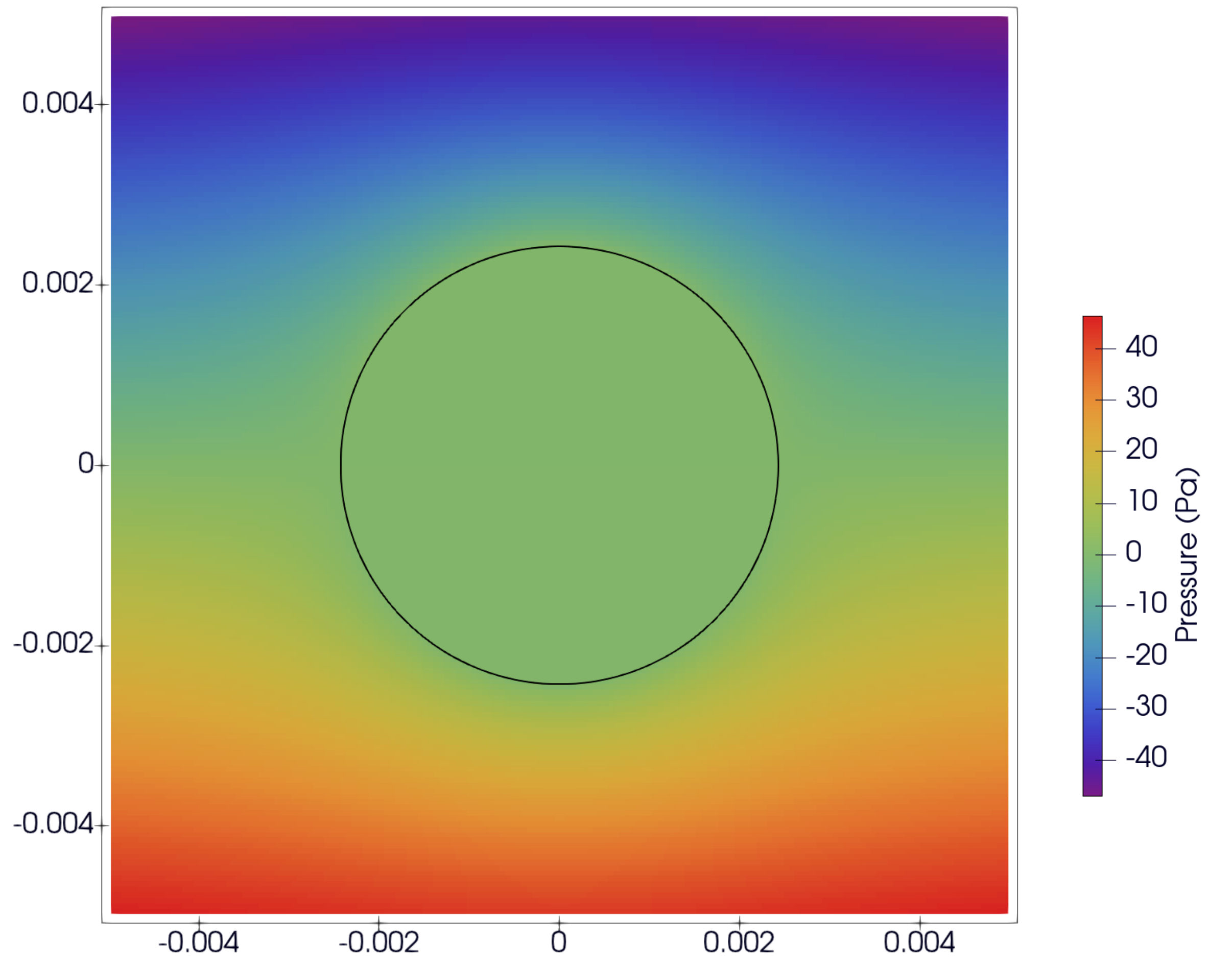}}%
    \put(0.45257185,-0.0120315){\color[rgb]{0,0,0}\makebox(0,0)[t]{\lineheight{13.25}\smash{\begin{tabular}[t]{c}$x_2$\end{tabular}}}}%
    \put(0.04802443,0.40352014){\color[rgb]{0,0,0}\rotatebox{90}{\makebox(0,0)[t]{\lineheight{13.25}\smash{\begin{tabular}[t]{c}$x_3$\end{tabular}}}}}%
  \end{picture}%
\endgroup%

}
  (a)
    \def\svgwidth{.45\textwidth}
    {\scriptsize
    \graphicspath{{figures/}}
\begingroup%
  \makeatletter%
  \providecommand\color[2][]{%
    \errmessage{(Inkscape) Color is used for the text in Inkscape, but the package 'color.sty' is not loaded}%
    \renewcommand\color[2][]{}%
  }%
  \providecommand\transparent[1]{%
    \errmessage{(Inkscape) Transparency is used (non-zero) for the text in Inkscape, but the package 'transparent.sty' is not loaded}%
    \renewcommand\transparent[1]{}%
  }%
  \providecommand\rotatebox[2]{#2}%
  \newcommand*\fsize{\dimexpr\f@size pt\relax}%
  \newcommand*\lineheight[1]{\fontsize{\fsize}{#1\fsize}\selectfont}%
  \ifx\svgwidth\undefined%
    \setlength{\unitlength}{1157.66927419bp}%
    \ifx\svgscale\undefined%
      \relax%
    \else%
      \setlength{\unitlength}{\unitlength * \real{\svgscale}}%
    \fi%
  \else%
    \setlength{\unitlength}{\svgwidth}%
  \fi%
  \global\let\svgwidth\undefined%
  \global\let\svgscale\undefined%
  \makeatother%
  \begin{picture}(1,0.78232127)%
    \lineheight{1}%
    \setlength\tabcolsep{0pt}%
    \put(0,0){\includegraphics[width=\unitlength,page=1]{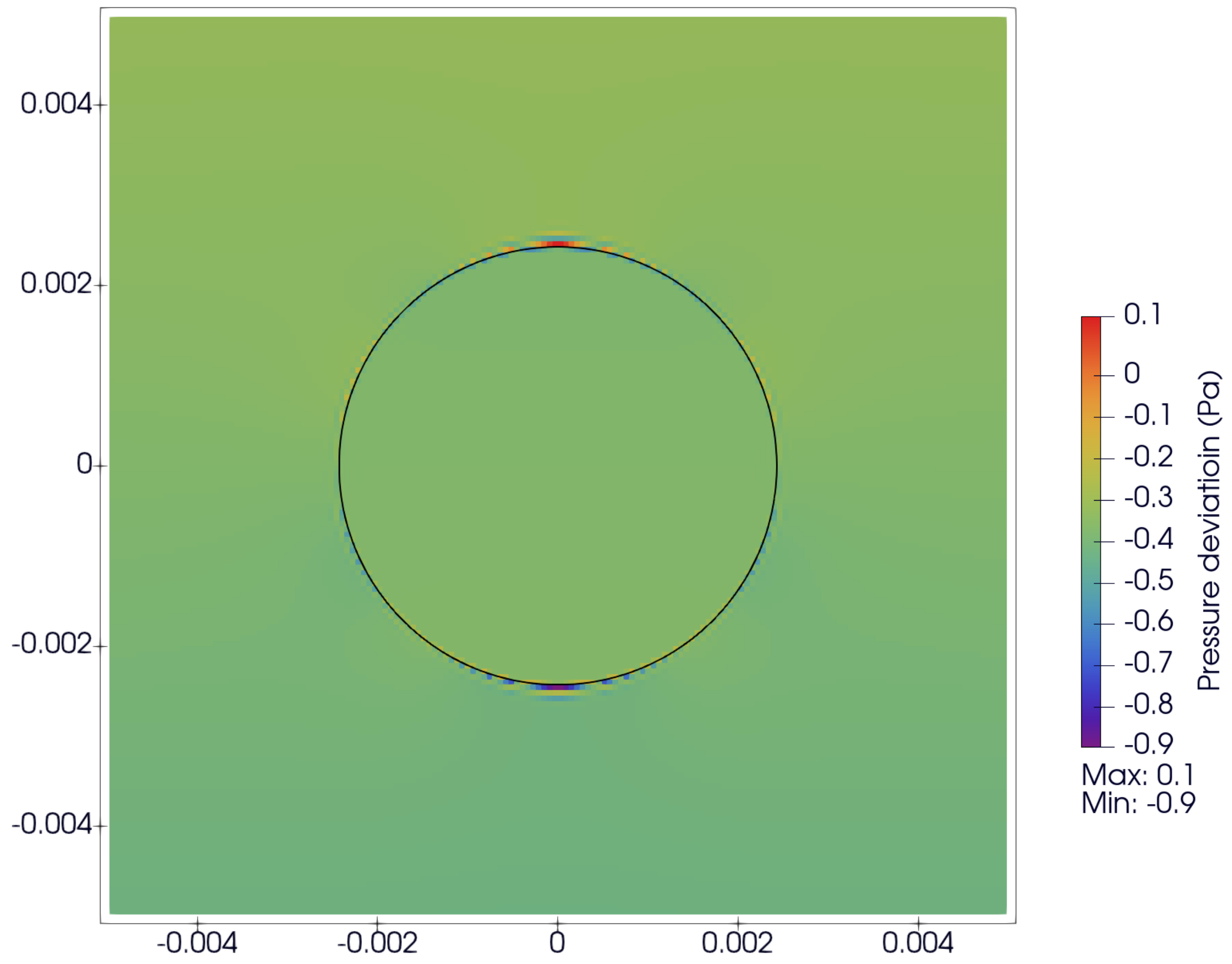}}%
    \put(0.45257185,-0.0120315){\color[rgb]{0,0,0}\makebox(0,0)[t]{\lineheight{13.25}\smash{\begin{tabular}[t]{c}$x_2$\end{tabular}}}}%
    \put(0.04802443,0.40352014){\color[rgb]{0,0,0}\rotatebox{90}{\makebox(0,0)[t]{\lineheight{13.25}\smash{\begin{tabular}[t]{c}$x_3$\end{tabular}}}}}%
  \end{picture}%
\endgroup%

}
(b)
  }
  \caption{Pressure distribution in the gas and liquid phases in the $x_2, x_3$-plane (at $x_1 = 0$) for a bubble accelerated from rest from the simulation at $t=2\times10^{-6}\,s$: (a) computed pressure field; (b) deviation of computed to analytical pressure. The black circle represents the bubble surface. Coordinate values in $m$.}
\label{fig:inipress-jun}
\end{figure}
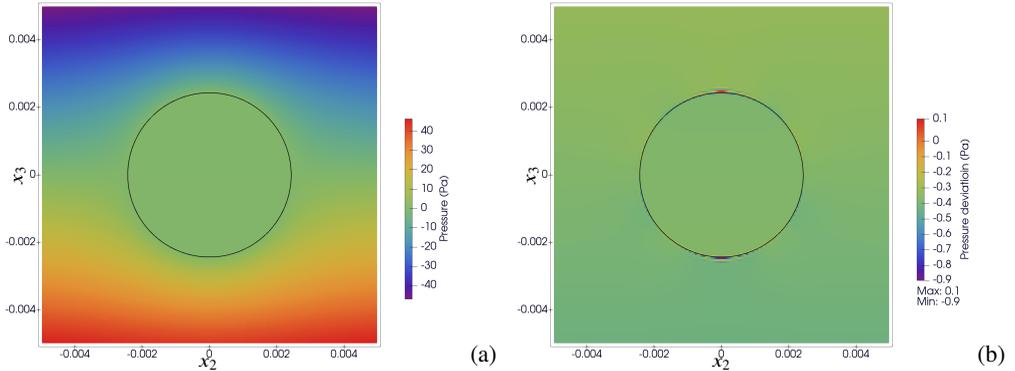
A large computational domain spanning $(-10R, 10R) \times (-10R, 10R) \times(-10R, 10R)$ was employed, with perfect slip conditions imposed on the lateral walls to ensure that any potential influence from the boundaries remains negligible. 
The mesh was statically refined to increase the accuracy around the bubble, while reducing computational costs (cf. Figure~\ref{fig:sketchnum}$\,$(b)). The initial coarse mesh has the resolution of $200\times200\times200$, giving the characteristic discretization length of $h=[\frac{D}{20}]$. The initial coarse mesh is then statically refined to obtain $(h=[\frac{D}{40},\, \frac{D}{80}])$ in the sub-region $(-2.5D, 2.5D)\times (-2.5D, 2.5D)\times(-2.5D, 2.5D)$ surrounding the bubble. 
 The densities of air $\rho^g$ and water $\rho^l$ correspond to their values under standard thermodynamic conditions, i.e., $ \rho^g = 1.225\,kg/m^3$, $\rho^l = 997\,kg/m^3$ respectively. The dynamic viscosity of air is $\mu^g = 0.0185\,mPa\cdot s$, while the viscosity of the liquid was set to $\mu^l = 30\,mPa\cdot s$. Since surface tension only shifts the pressure inside the bubble by a constant value, it is not relevant for the initial acceleration and has been set to zero. Figure~\ref{fig:sketchnum}$\,$(a) gives also the boundary conditions for dynamic pressure and velocity. The dynamic pressure $p_d = p - \rho(\gbf\cdot\xbf)$ is used in solving the two-phase Navier-stakes equations, with the blended density $\rho = \alpha \rho^l + (1-\alpha)\rho^g$. 
 
 In this study, the bubble's center position $\xbf_B$ and velocity $\vbf_B$ are estimated by summing the cell values weighted by cell's gas volume $(1-\alpha_c)V_c$, as represented by
 \begin{equation}
     \xbf_B = \dfrac{\sum_{c\in N_c}\xbf_c(1-\alpha_c)V_c}{\sum_{c\in N_c}(1-\alpha_c)V_c} \quad \text{ and } \quad
     \vbf_B = \dfrac{\sum_{c\in N_c}\vbf_c(1-\alpha_c)V_c}{\sum_{c\in N_c}(1-\alpha_c)V_c},
 \end{equation}
respectively, where $c$ denotes the mesh cells, $N_c$ is the set of cells within the computational domain, $\xbf_c$ is the cell center position, and $\vbf_c$ the computed cell velocity. The acceleration of the rising bubble at the time step $n+1$ is then calculated as 
\begin{equation}
\abf_B^{n+1} = \frac{\vbf_B^{n+1}-\vbf_B^{n}}{t^{n+1} - t^n}.
\end{equation}
Since the bubble rises in the $x_3$-direction, the $x_3$-component of $\xbf_B$, $\vbf_B$ and $\abf_B$ is adopted to analyze the simulation results. In depicting the numerical results, the initial acceleration is normalized by $g$.
  
\begin{figure}
\centering
\includegraphics[width=.55\textwidth]{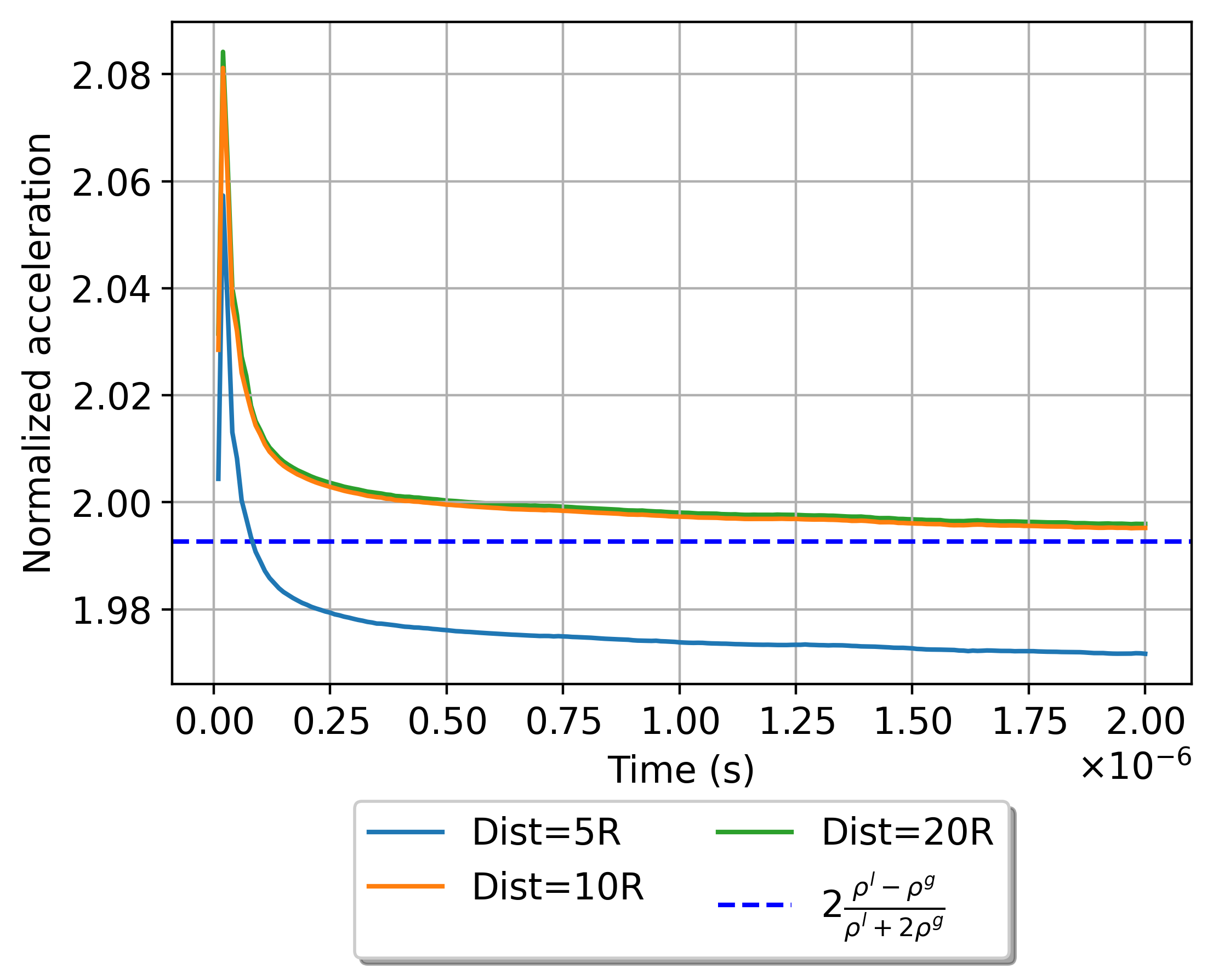}
  \caption{Effect of the wall distance to the bubble center on the initial acceleration: Dist $=[5R,\,10R,\,20R]$.}
\label{fig:WallDistStudy}
\end{figure}
   
An essential assumption in Section~\ref{sec:theory} is that the bubble is placed in an infinite region to avoid wall effects. To approximate this in the numerical simulations, the computational domain must be sufficiently large. We tested three domains with varying distances between the walls and the bubble center: Dist $=[5R,\,10R,\,20R]$. The resolution and time step size were kept identical across these cases.  As shown in Figure~\ref{fig:WallDistStudy}, the domain size has negligible impact on the stable acceleration when the domain is sufficiently large (Dist $\ge$ 10R). In contrast, the results for the case with Dist = 5R are significantly lower than both the analytical result and the other simulations, indicating a noticeable influence from the walls. The domain size of $(-10R, 10R)\times (-10R, 10R)\times(-10R, 10R)$ was selected to balance computational efficiency with accuracy. 

\begin{figure}      
\includegraphics[width=.455\textwidth]{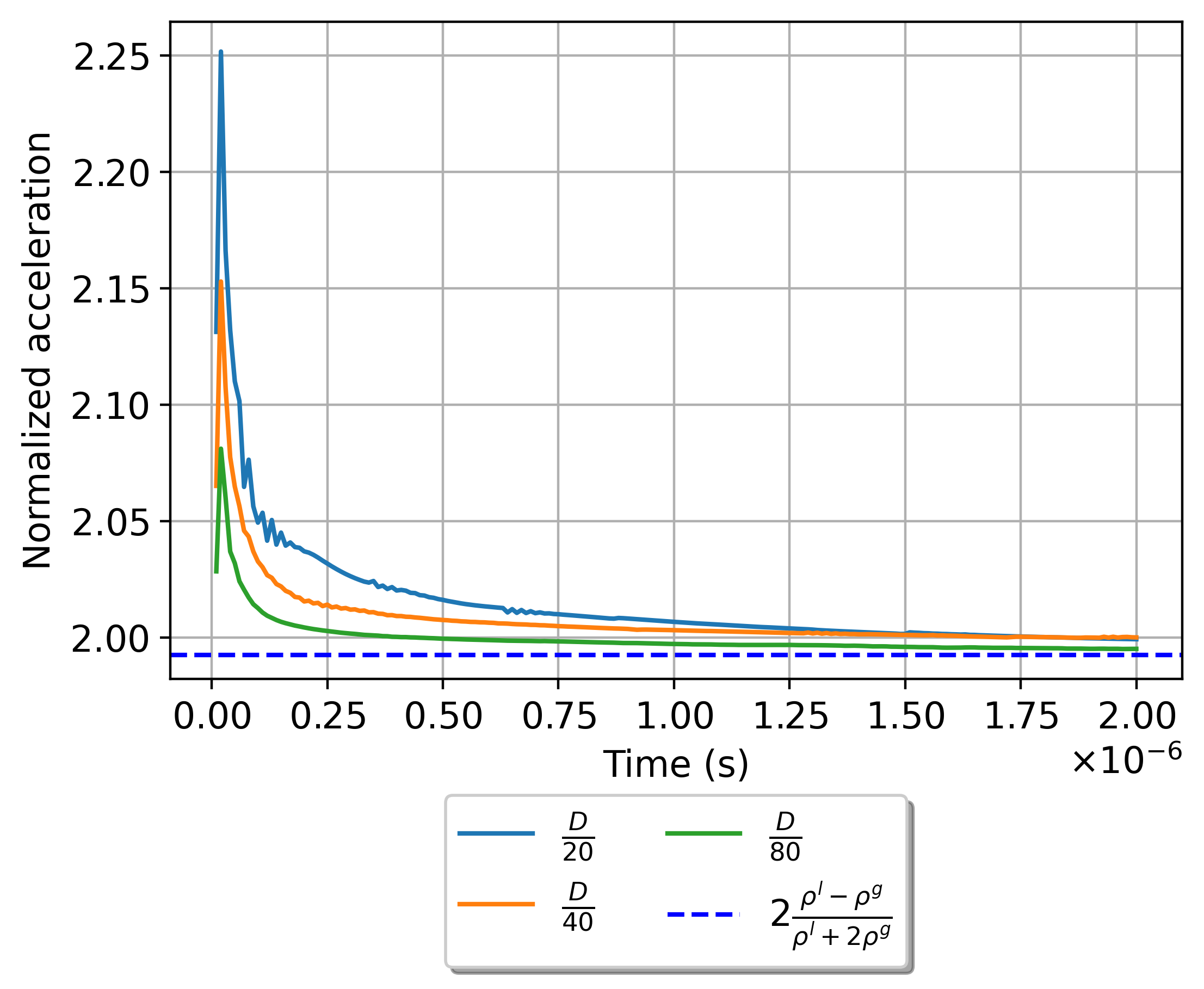} (a)
\includegraphics[width=.455\textwidth]{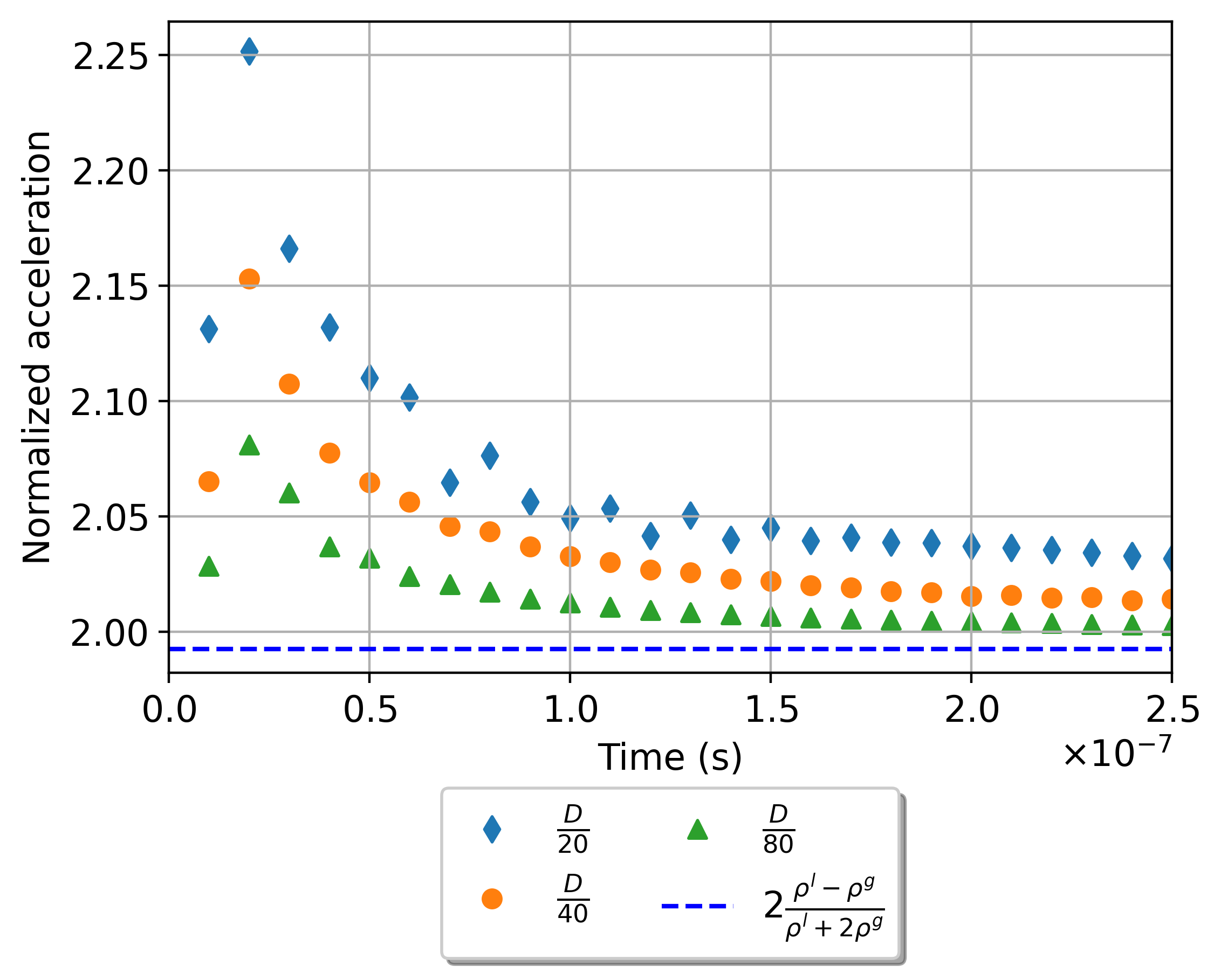} (b)
  \caption{Mesh convergence tests based on the initial acceleration (normalized by $g$) with a fixed time step size $\Delta t=10^{-8}\,s$: (a) full view; (b) results in the first 25 time steps.}
\label{fig:MeshConvergence}
\end{figure}
A mesh convergence test was conducted with three grid resolutions: $h=[\frac{D}{20}, \frac{D}{40},\frac{D}{80}]$, where the bubble diameter $D$ is resolved by $20$, $40$ and $80$ cells, respectively. As depicted in Figure~\ref{fig:MeshConvergence}$\,$(a), increasing the resolution reduces numerical oscillations and brings the computed acceleration closer to the analytical value. All acceleration plots have an initial impulse at the start of the simulation. A zoomed-in view of the first 25 time steps is provided Figure~\ref{fig:MeshConvergence}$\,$(b). The numerically computed acceleration does not match the analytical acceleration in these first time steps because the initial PLIC interface requires more time to align itself with the forces approximated by the Finite Volume discretization. The highest resolution, i.e., $\frac{D}{80}$, was used throughout unless otherwise specified. 

\begin{figure}      
\centering
\includegraphics[width=.5\textwidth]{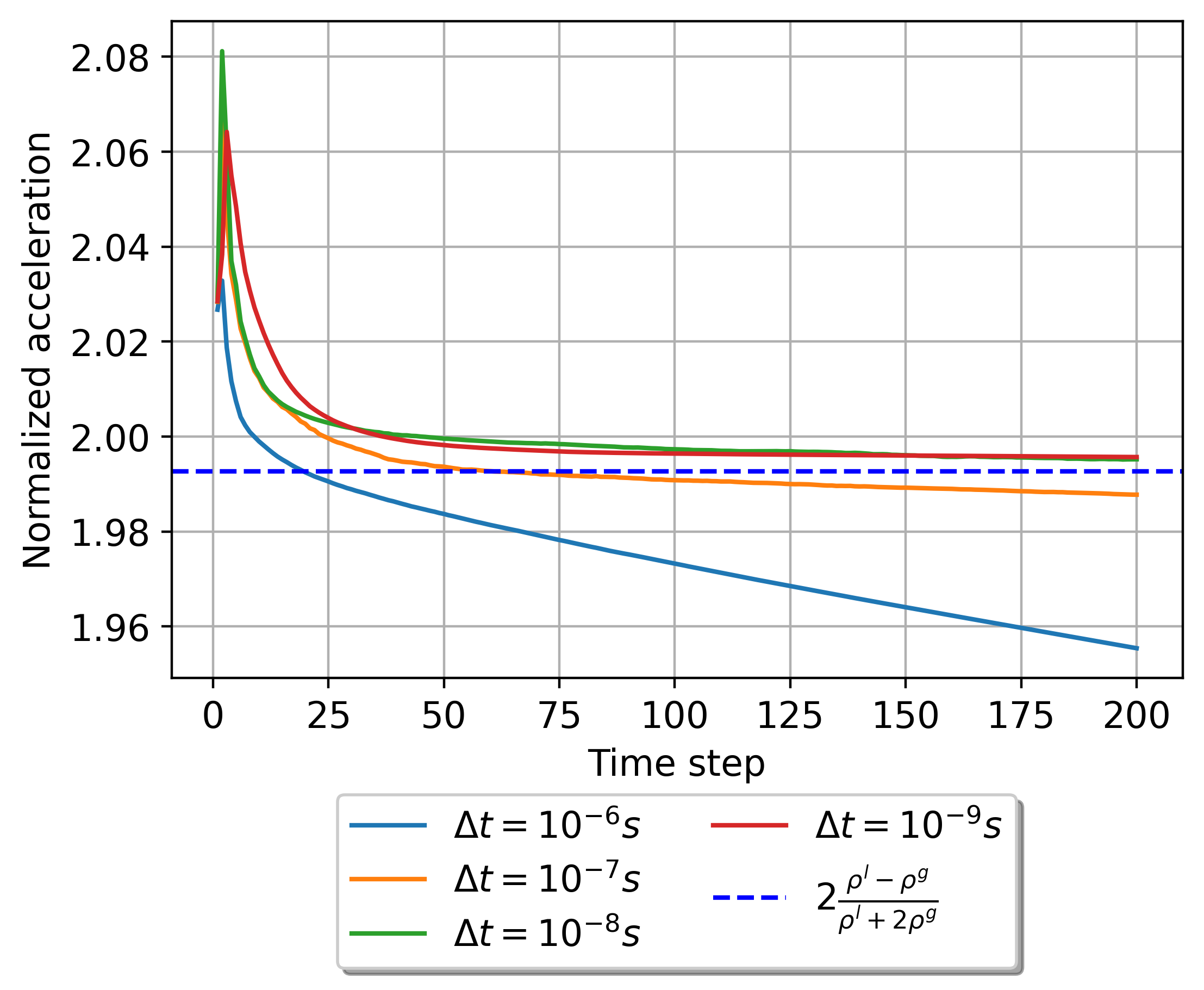}
  \caption{Simulated initial acceleration (normalized by $g$). Time step size convergence test.}
\label{fig:DeltatConvergence}
\end{figure}
To capture the acceleration as $t\to 0+$, a very small time step is required. We tested the 4 different time step sizes \ $\Delta t =[10^{-6},10^{-7},10^{-8},10^{-9}]s$ for the first $200$ time steps.
The initial acceleration cannot be stably captured by using a coarse time step size, e.g.\ $\Delta t= 10^{-6}s$, as shown in Figure~\ref{fig:DeltatConvergence}~(a). In contrast, we can obtain the stable initial acceleration through a smaller time step, such as $\Delta t= 10^{-8}s$ or $10^{-9}s$. For subsequent simulations, we chose $\Delta t= 10^{-8}s$. 

\begin{figure} 
    \centering
    \centerline{\includegraphics[width=0.5\linewidth]{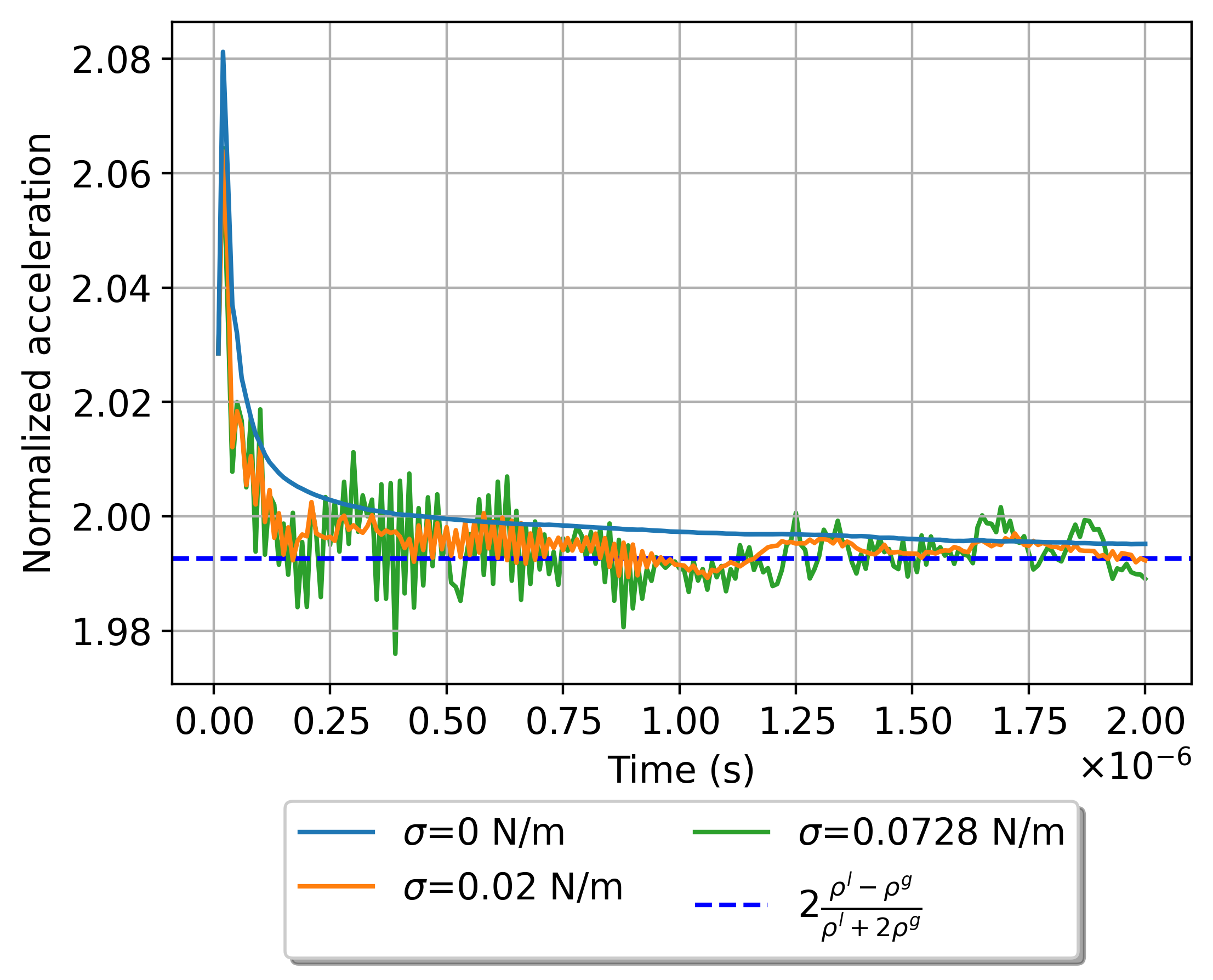}(a)
    \includegraphics[width=0.47\linewidth]{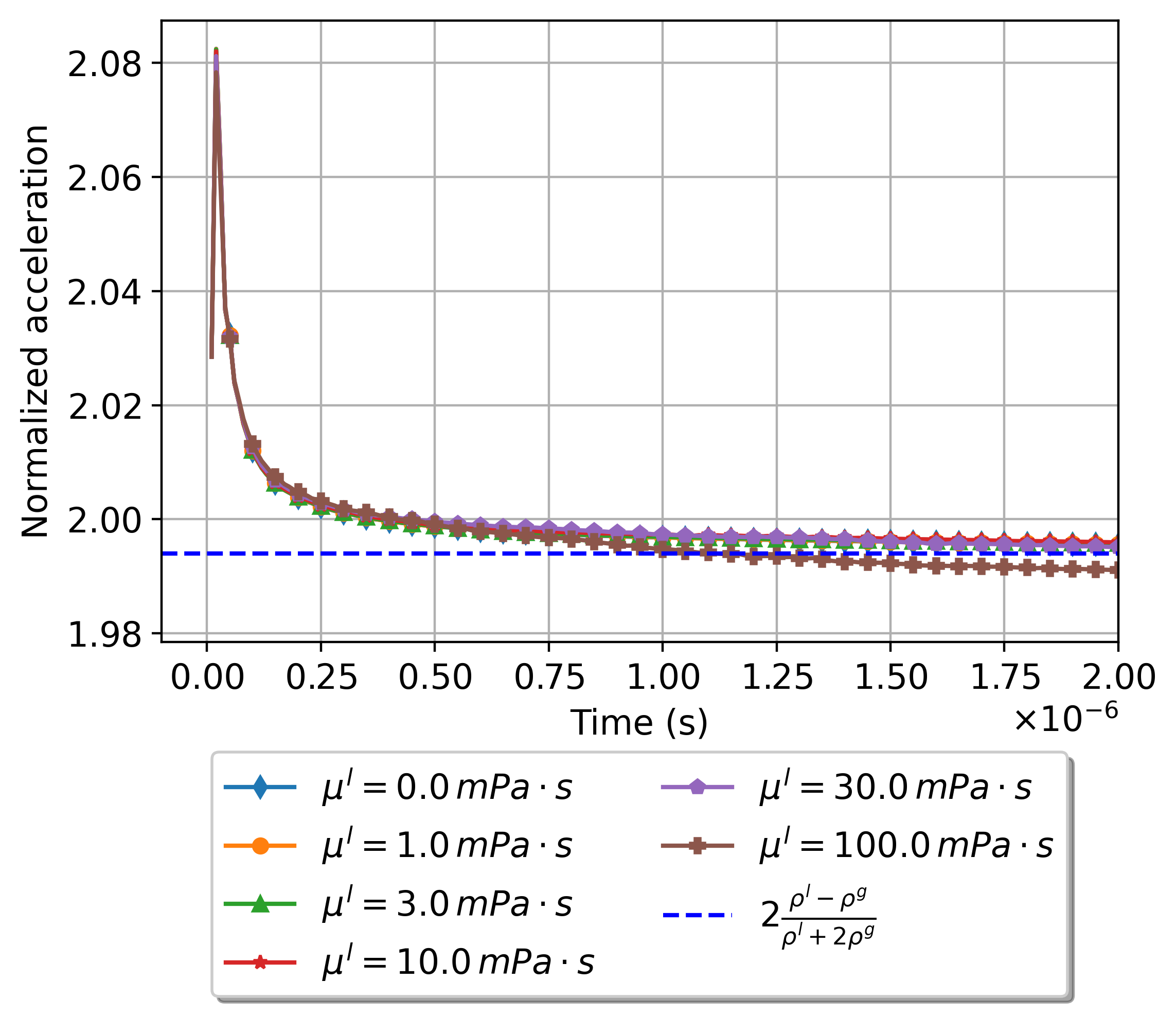}(b)}
    \caption{Simulated initial acceleration (normalized by $g$). (a) for surface tension $\sigma=[0.0,\, 0.02,\, 0.0728]\,N/m$; (b) for dynamic liquid viscosities $\mu^l=[0.0,\, 1.0,\, 3.0,\, 10.0,\, 30.0,\, 100.0]\,mPa\cdot s$.}
    \label{fig:STEffect}
\end{figure}
First, the effects of surface tension and viscosity were investigated by simulating the rising bubble with different surface tension coefficients $\sigma=[0.0, 0.02, 0.0728]\,N/m$, and different dynamic viscosities of the liquid $\mu^l=[0.0,\, 1.0,\, 3.0,\, 10.0,\, 30.0,\, 100.0]\,mPa\cdot s$. Incorporating surface tension introduced numerical noise, as illustrated by Figure~\ref{fig:STEffect}$\,$(a), which is caused by the known difficulty of VOF methods in accurately approximating the surface tension force. Despite these disturbances, the acceleration plots for the cases with non-zero surface tension remain relatively stable and align closely with the analytical solution. Nevertheless, to avoid parasitic currents, surface tension was switched off in all further simulations.
The plots in Figure~\ref{fig:STEffect}$\,$(b) display the viscous effects on the initial acceleration. In this very early phase of the bubble's rise, the viscous drag is visible only for the highest viscosity, being 100 times higher than water ($\mu^l=100.0\,mPa\cdot s$). Note that the bubble's rise speed $U_B=|\vbf_B|$ for very small times $t$ is approximately $U_B=2gt$, and this velocity induces viscous forces of about $\eta U_B R= 2\eta g t R$.
Hence the ratio between the viscous force and the (total) accelerating force, i.e.\ $2\rho^g V_B g$, has the magnitude $\eta t / (\rho^g R^2)$. Therefore, the viscous force is negligible as long as
$t\ll \frac{\rho^g}{\rho^l} t_{\rm visc}$, where $t_{\rm visc}=R^2/\nu^l$ is the viscous time scale, i.e.\ the characteristic time for viscous momentum diffusion across a distance $R$.
At the time instance $t = 1 \mu s$, we have $U_B=2 \times 10^{-5} m/s$ and a viscous force of magnitude $10^{-9} N$, to be compared to an accelerating (total) force of $10^{-6} N$.
This gives an order of magnitude estimate, supporting the fact that viscous effects after $1 \mu s$ are negligible for liquid viscosities below $\mu^l=100.0\,mPa\cdot s$, but might become visible at about this value of $\mu^l$.

At the time instant $t = 2\times 10^{-6} s$, the simulation yields the pressure field in the flow as seen in Figure~\ref{fig:inipress-jun}$\,$(a). The data in Figure~\ref{fig:inipress} were obtained from the theoretical calculation, while those in Figure~\ref{fig:inipress-jun}$\,$(a) were computed by simulations. The pressure from the simulation is very close to the analytical result of Figure~\ref{fig:inipress} from Section~\ref{sec:theory}. We report the accuracy using the difference between the analytical pressure and the pressure from the simulation in Figure~\ref{fig:inipress-jun}$\,$(b).  The maximum absolute deviation of the computed pressure from the theoretical calculation is less than $1\,Pa$, which demonstrates excellent agreement with the analytical solution.

\begin{figure}      
  \centerline{\includegraphics[width=0.5\textwidth]{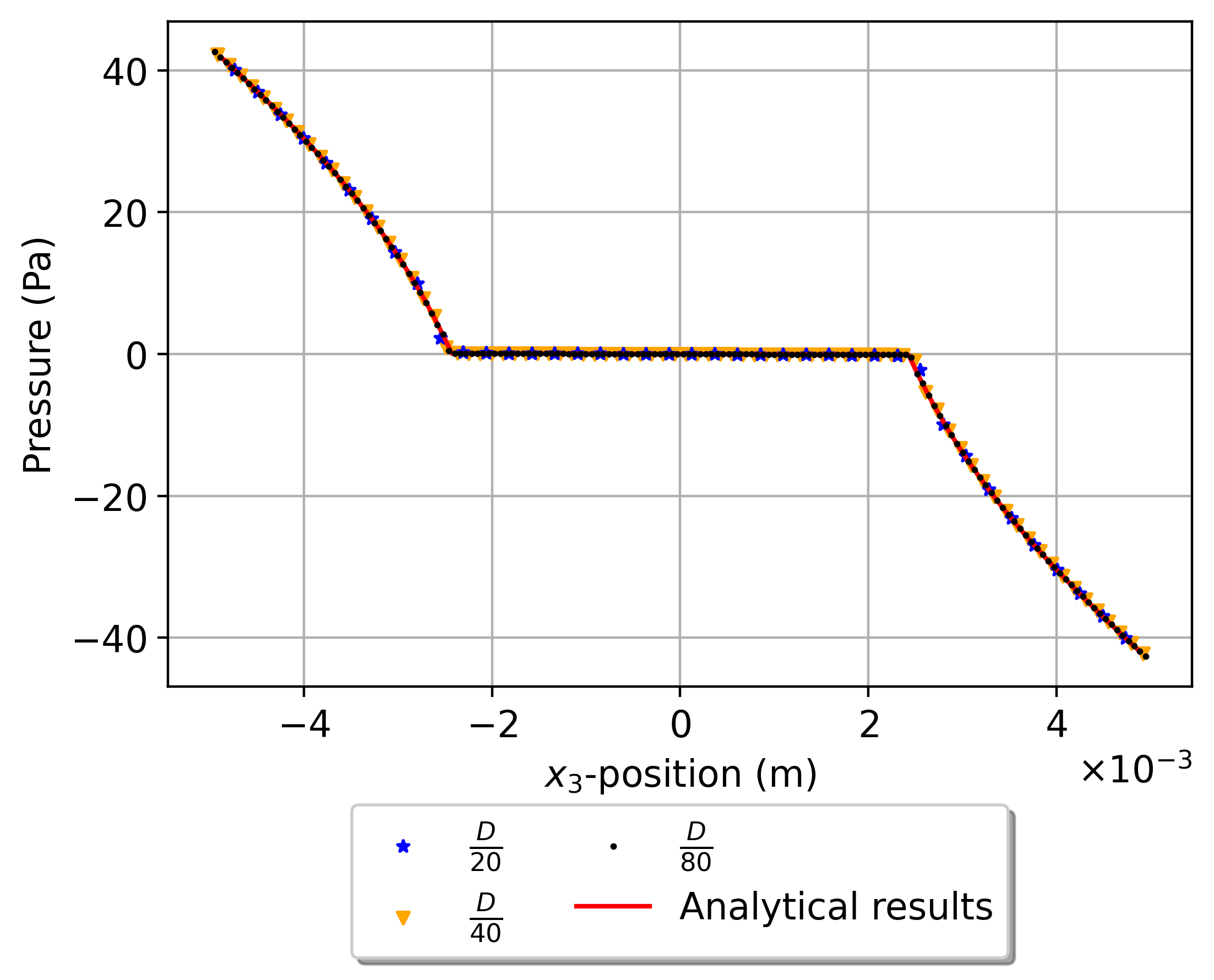}(a)
  \includegraphics[width=0.5\textwidth]{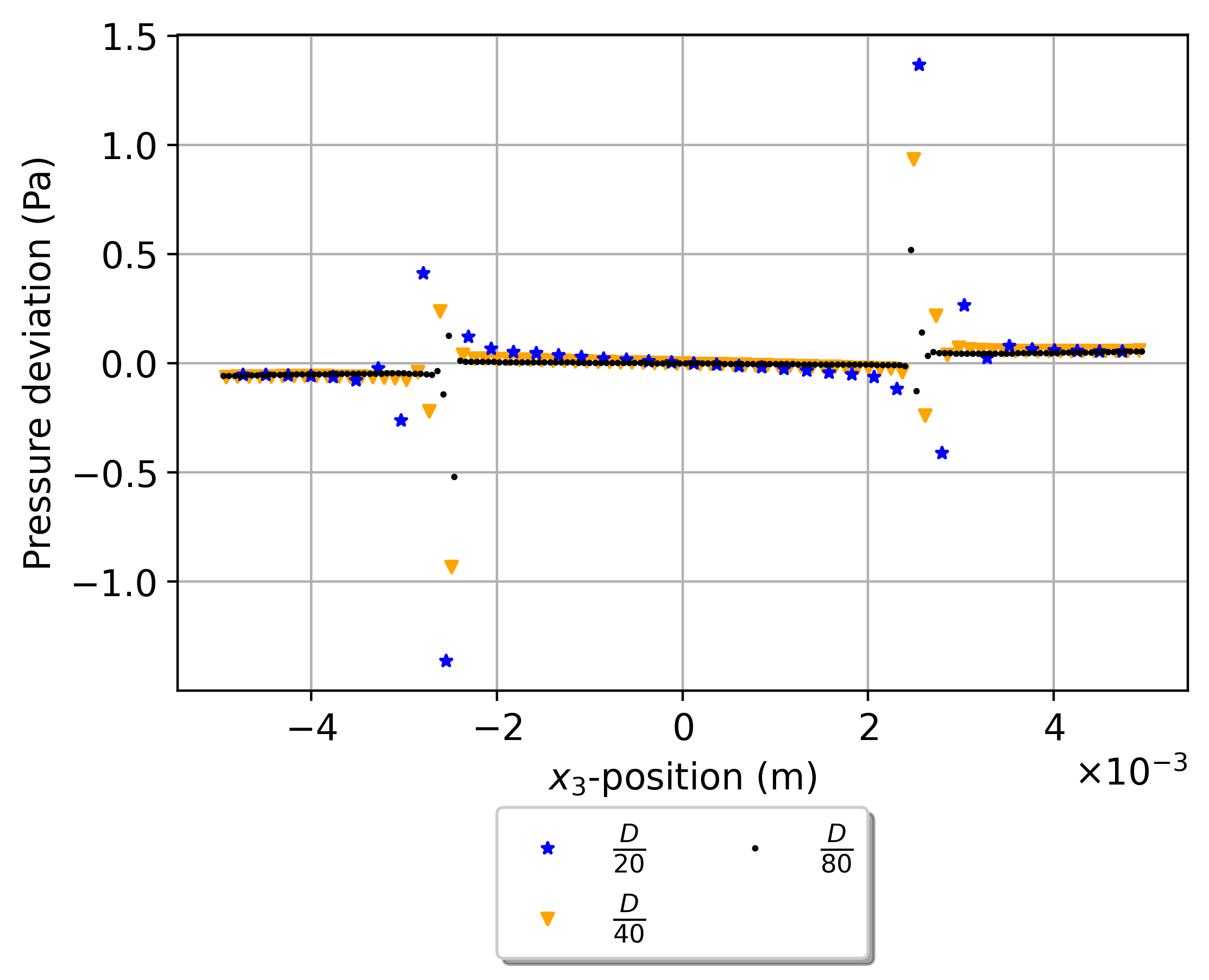}(b)}  
  \centerline{\includegraphics[width=0.5\textwidth]{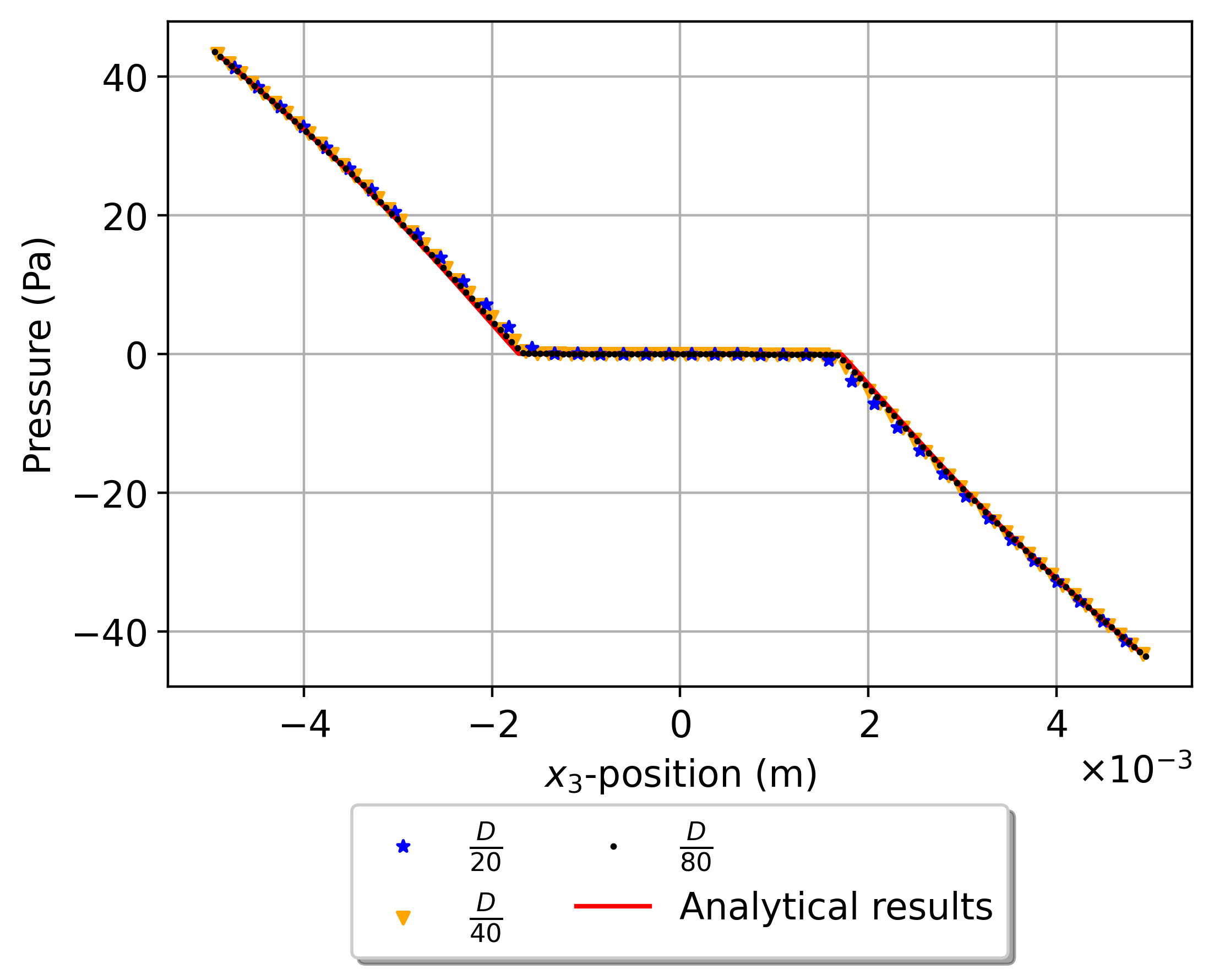}(c)
  \includegraphics[width=0.5\textwidth]{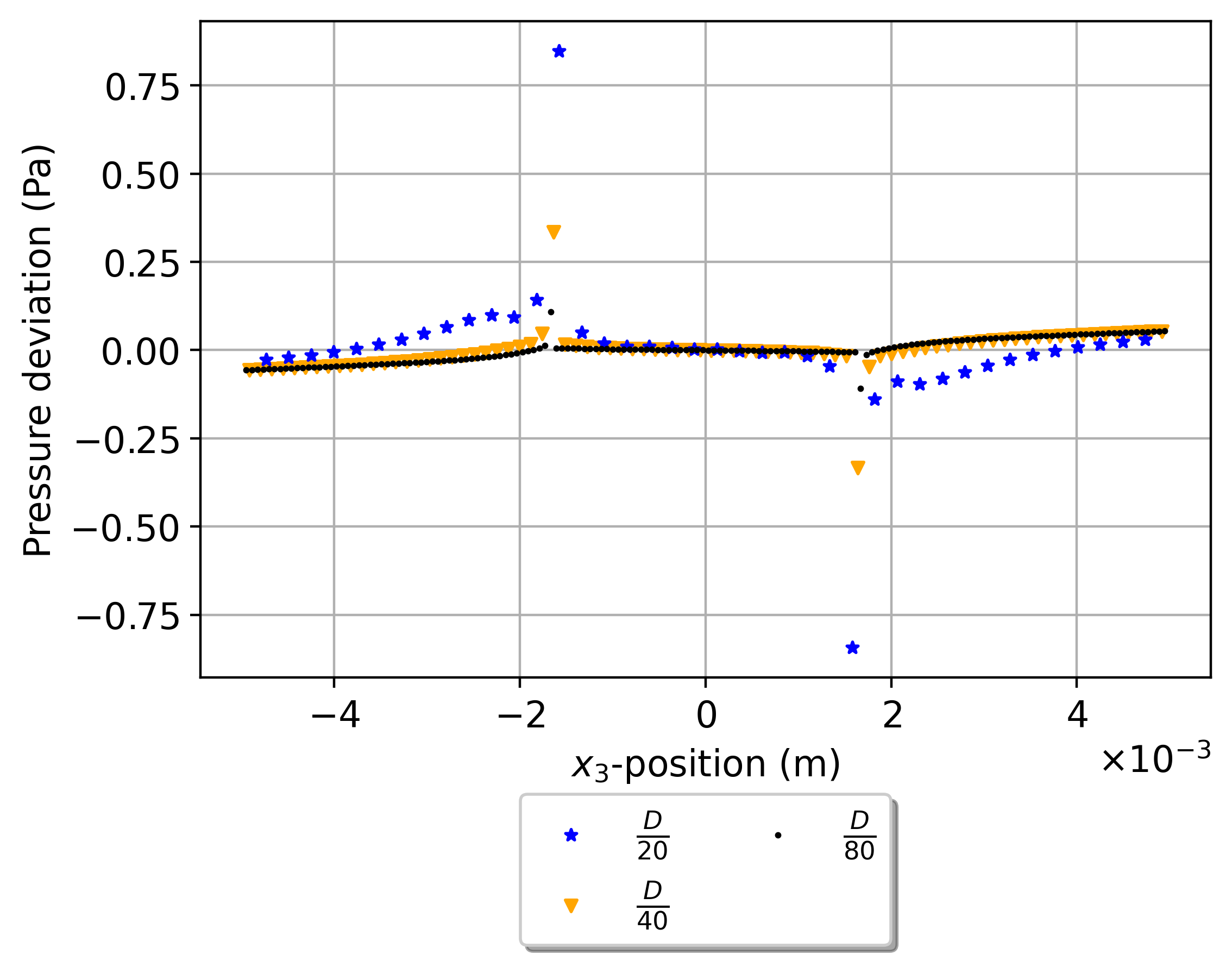}(d)}
  \caption{Pressure profiles along the $x_3$-axis (upper row) and 
  along a parallel vertical line crossing the point $(\frac{R}{2}, \frac{R}{2}, 0)$ (lower row). Comparison of the analytical solution and the numerical results at the time instant $t=2\times 10^{-6} s$, with resolutions of 20, 40 and 80 grid cells for the bubble diameter. (a), (c) Profiles displaying the liquid pressure up to one bubble radius above and below the bubble; (b), (d) zoom-in view, displaying the pressure deviation from the analytical solution. }
\label{fig:pressprofiles}
\end{figure}

A comparison between the analytical and the numerical solutions is shown in Figure~\ref{fig:pressprofiles}, where pressure profiles are sampled along the $x_3$-axis and along another vertical line passing through the bubble. The simulations were carried out with resolutions of $20$, $40$ and $80$ cells per bubble diameter. Figures~\ref{fig:pressprofiles}$\,$(a), (c) show the pressure up to one bubble radius outside the bubble, while Figures~\ref{fig:pressprofiles}$\,$(b) and (d) display the pressure deviation between simulations and analytical results. On the large range of pressure values shown in Figures~\ref{fig:pressprofiles}$\,$(a) and (c) for the two different samplings, agreement between the simulations and the analytical solution appears excellent. The zoomed-in views in Figures~\ref{fig:pressprofiles}$\,$(b) and (d) reveal minor discrepancies between the solutions. 
Except for a thin layer around the interface, the pressure deviations between the numerical and analytical results are very small (about $\pm\,0.05 \,Pa \, s$) and nearly  constant, except for the coarsest mesh. 
Larger deviations appear near the interface, where $0 < \alpha_c < 1$. Refining the mesh reduces these deviations as shown in Figure~\ref{fig:pressprofiles}$\,$(b) and (d), thus significantly increasing the accuracy.  
These deviations are caused by the numerical treatment of the body force due to gravity. Due to the jump in the mass densities at the interface, the gravity force has a jump, too. The numerical treatment of the pressure Poisson equation also employs a modified pressure. Similar to what is done in the analytical calculation in Section~\ref{sec:theory} above, a static pressure component  is subtracted from the total pressure, which is build with the local, i.e.\ discontinuous at the bubble surface, mass density. The Poisson equation for the resulting dynamic pressure in the one-field formulation contains the divergence of  $(\mathbf{g}\cdot\mathbf{x})\nabla \rho$ on the right-side. The discretisation of this introduces some smearing such that this term  becomes visible in the numerical solution close to the interface. This also explains that there is a single oscillation appearing at the interface positions. 
\begin{figure}      
  \centerline{\includegraphics[width=0.5\textwidth]{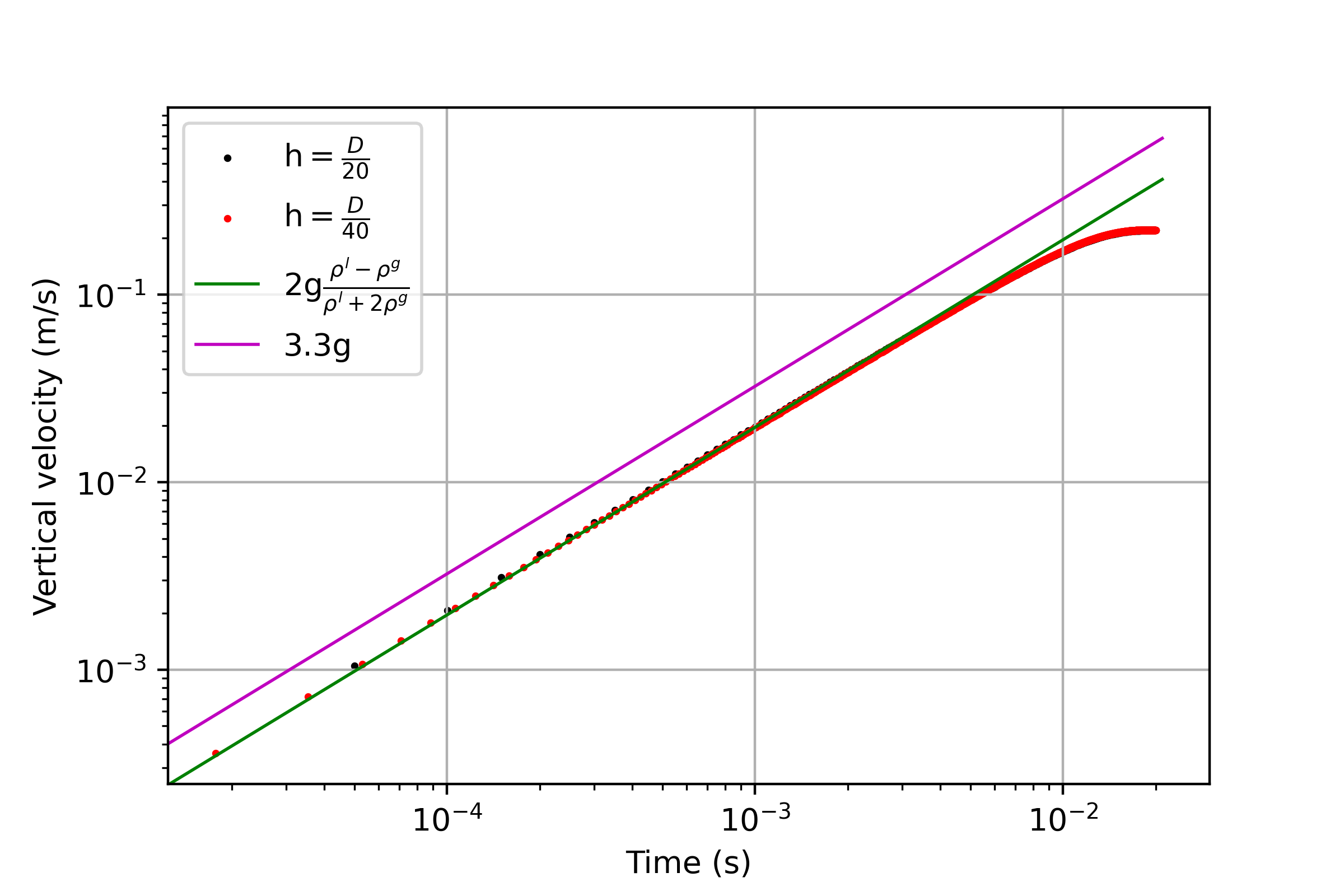}(a)
  \includegraphics[width=0.5\textwidth]{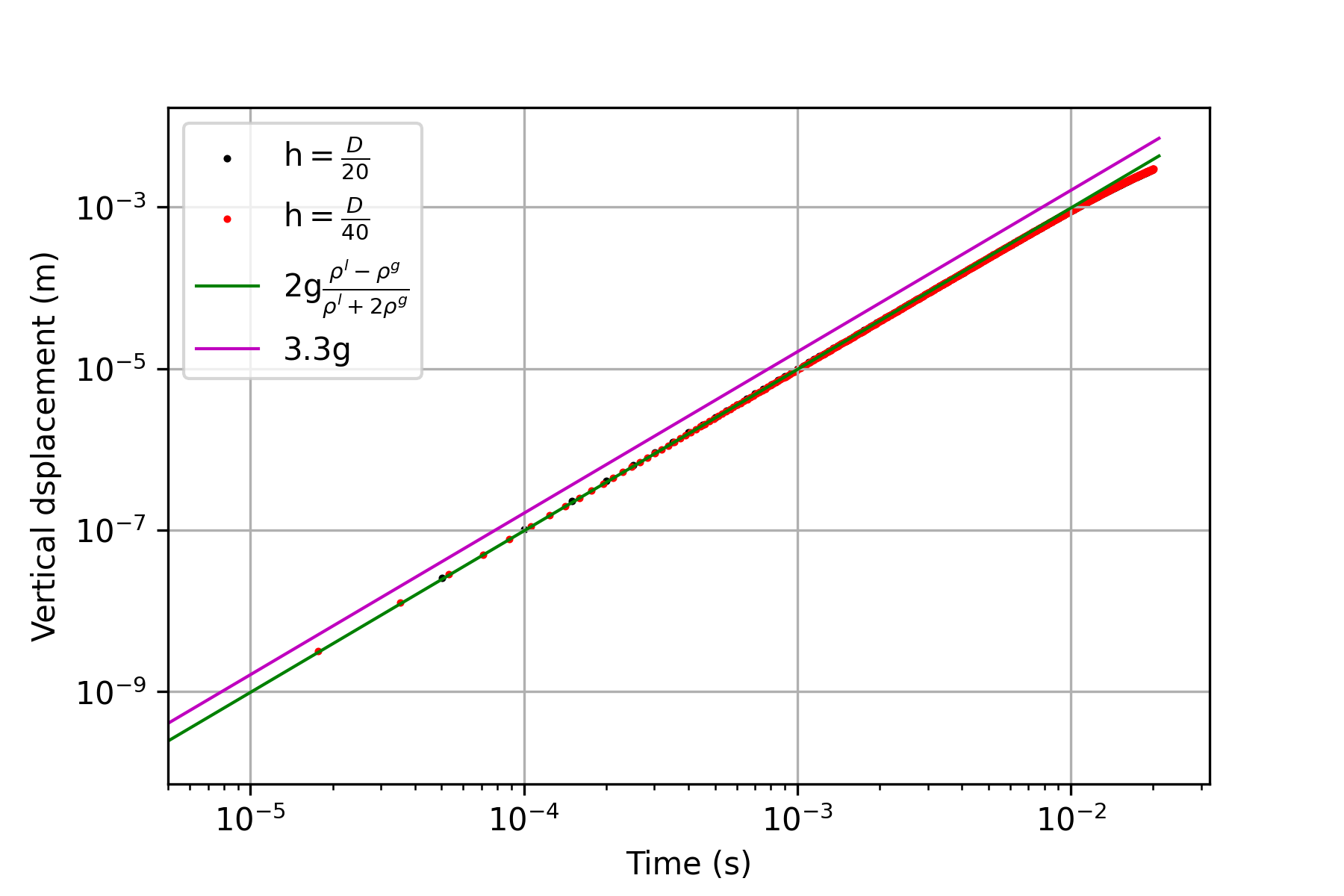}(b)} 
  \caption{(a) The bubble velocity and (b) the bubble centroid displacement from its initial position, both in the $x_3$ direction, as functions of time. Comparison of the numerical simulations at two different mesh resolutions (dots) with the analytical solution (green lines) and the line corresponding to an initial acceleration of $3.3g$ (purple line), as found in \cite{Dominik-Cassel-2015}.}
\label{fig:velocitydisplace}
\end{figure}
\begin{figure}              
    \centerline{\includegraphics[width=0.5\linewidth]{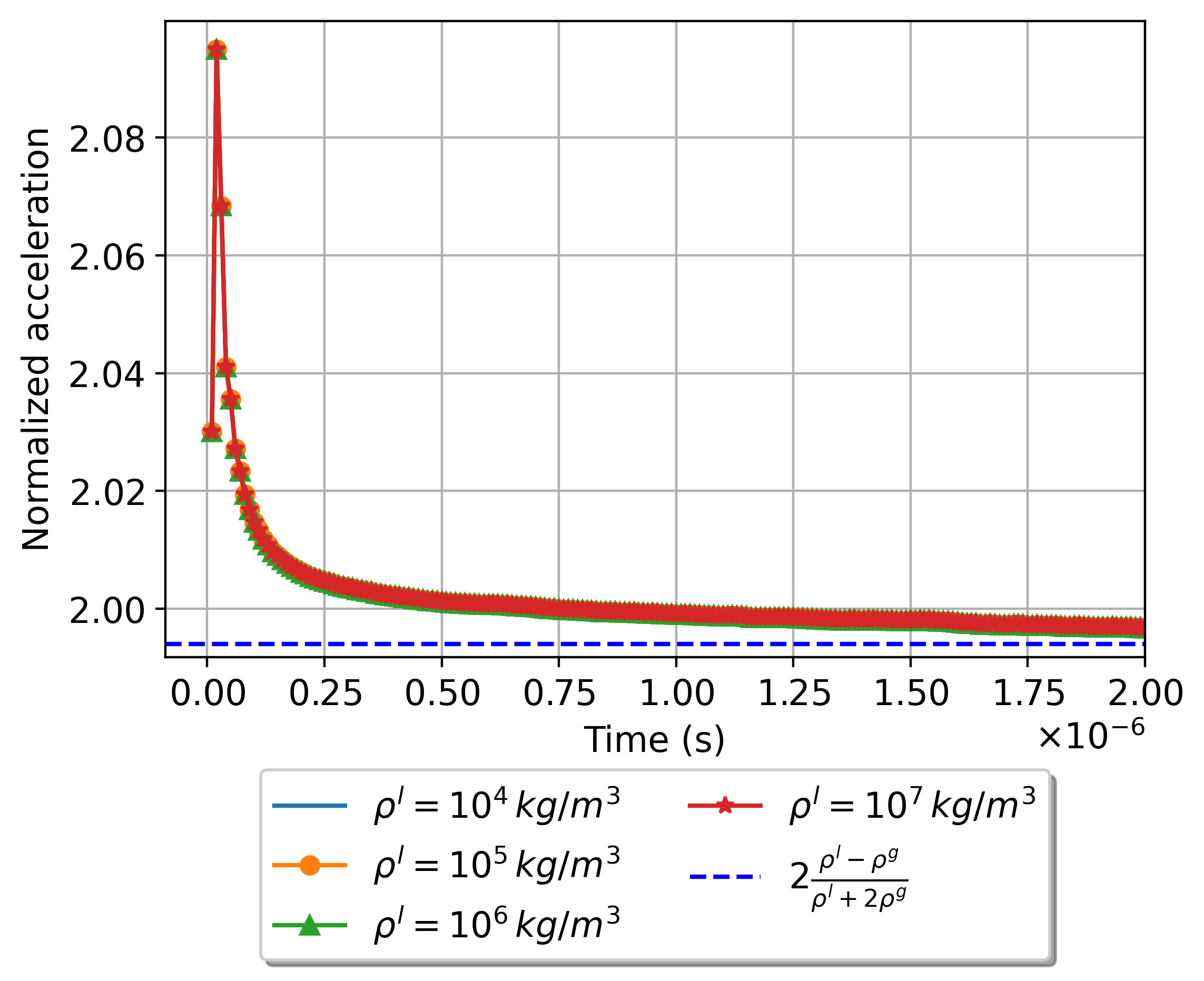}(a)
    \includegraphics[width=0.5\linewidth]{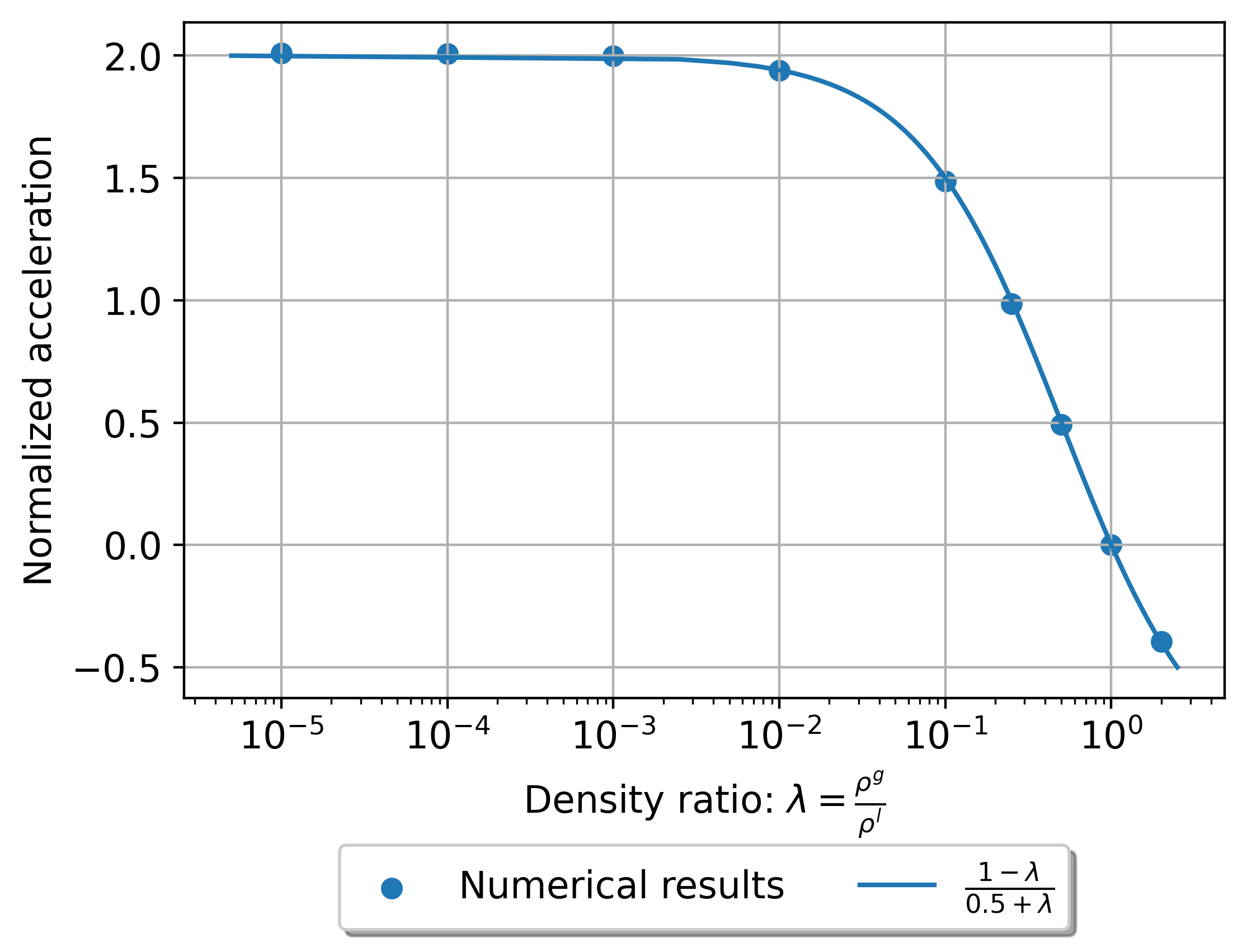}(b)}
    
    \caption{Numerically simulated initial acceleration (normalized by $g$) for different material pairings: (a) gas/liquid pairings with the same density ratio of air/water and different liquid densities $\rho^l=[10^4, 10^5,10^6,10^7]kg/m^3$; (b) gas/liquid pairings with the same liquid density as water and different density ratios $\lambda=\frac{\rho^g}{\rho^l}=[10^{-5}, 10^{-4}, 10^{-3}, 10^{-2}, 10^{-1}, 0.25, 0.5, 1.0, 2.0]$.}
    \label{fig:IniAcc_pairings}
\end{figure}

The motion of the bubble is compared to results from  \cite{Dominik-Cassel-2015} in Figure~\ref{fig:velocitydisplace}. The bubble velocity in Figure~\ref{fig:velocitydisplace}$\,$(a) increases at the constant acceleration of approximately $2g$. The corresponding bubble centroid displacement is shown in Figure~\ref{fig:velocitydisplace}$\,$(b). Our results demonstrate that within the first $5\times 10^{-2}\,s$, the simulated bubble velocity and the displacement of the bubble agree very well with lines which correspond to an acceleration of $2g$. This is in contrast to the results by \cite{Dominik-Cassel-2015}. 
We have additionally simulated initial accelerations for different gas/liquid pairings. When the density ratio is held constant, varying the liquid density does not impact the initial acceleration, as shown in Figure~\ref{fig:IniAcc_pairings}. Stable accelerations at $t=2\times10^{-6}s$ with $\Delta t = 10^{-8}s$ were obtained for 9 gas/liquid pairings. 
Rewriting the initial acceleration from (\ref{eq:ini-acc}) as a function of the density ratio 
$\lambda=\rho^g /\rho^l$, we obtain the expression $\abf_B = (1-\lambda)/(0.5+\lambda)$. The excellent agreement between this analytical solution and the numerical simulations in Figure~\ref{fig:IniAcc_pairings}$\,$(b) demonstrates the  applicability of this model across strongly varying density ratios.

\section{Discussion and Conclusions}\label{sec:discussconclus}
The classical studies of the initial acceleration of spherical fluid particles either use potential flow theory, assuming irrotational flow inside and outside the particle, or apply the Stokesian stream function, assuming axial symmetry and vanishing swirl. Due to a dimensional analysis performed by Mougin and Magnaudet in the context of solid particles, see Appendix~A in \cite{mougin2002generalized}, it is to be expected that the development of rotational flow components does not affect the added mass, hence the initial acceleration. However, while this makes it very plausible that the classical result holds true without additional assumptions, except possibly a fixed shape, it does not provide a rigorous proof. Triggered by the surprising numerical results reported in \cite{Dominik-Cassel-2015}, showing numerical results of accelerating bubbles with initial spherical shape and an acceleration of $3.3g$, we revisited this classical question. 

We exploit the fact that the flow fields inside and outside of the bubble initially vanish for this particular problem. This allows to reduce the two-phase Navier-Stokes equations with capillary interface to a transmission problem for the pressure in the two phases. This can be solved analytically for spherical geometry, and this way we obtain the classical result for the initial acceleration, but this time without any additional assumption. A main step here was to be able to allow for an immediate change of the shape, evolving from the initial spherical one. For this, it is important to not use the momentum balance in the form (\ref{E9}), i.e.\ for $\rho \vbf$, but to first divide by the phase-specific mass density, before a jump condition for the normal derivative of the pressure at the interface is derived. The result remains valid for any liquid/fluid pairing.

It is worthwhile to think a bit more about the resulting value of about $2g$ for a spherical air bubble accelerating from rest inside an ambient liquid phase. A naive approach would be to apply the Archimedian principle to the bubble as an extended fluid phase -- instead of a point mass, for which a surrogate model is to be built. In this case, one would obtain an upward force being about 1000 times larger than the gravitational force acting downwards on the bubble. Note that this would imply the undisturbed hydrostatic pressure in the liquid up to the bubble's surface. This, in turn, would obviously lead to an initial acceleration that is unrealistically large. Consequently, this shows that the Archimedian principle does not apply to an accelerating bubble. This is indeed true, and it was known to Archimedes -- at least this is what we can assume from what is known as his Proposition 6 in his book 'On Floating Bodies' (see p.\ 257 in \cite{arquimedes1897works}):\vspace{6pt}
\begin{center}
    \emph{If a solid lighter than a fluid be forcibly immersed in it, the solid will be driven upwards by a force equal to the difference between its weight and the weight of the fluid displaced.}\vspace{6pt}
\end{center}
In other words, if we would be able to keep the bubble from rising by keeping it in a state of force equilibrium, the situation would be fundamentally different from the accelerating one. As a thought experiment, we can think of capturing the air inside a thin shell, such as a 'ping-pong ball', and keep the shell from rising by fixing it to a thread attached to the bottom of the container filled with water. In this scenario, the pressure outside the 'ping-pong ball' would be the undisturbed hydrostatic pressure in the water, and the force acting on the thread would be the difference between the Archimedian buoyancy force and the weight of the 'ping-pong ball'. If the thread is cut, however, the pressure will instantaneously change from the hydrostatic to the pressure field that is shown in figure~\ref{fig:inipress}. Recall that we assume incompressibility, hence infinite sound speed inside the fluids. Therefore, the change in the pressure profile would be instantaneous.
\begin{figure}     
    \centering
    \includegraphics[width=0.6\linewidth]{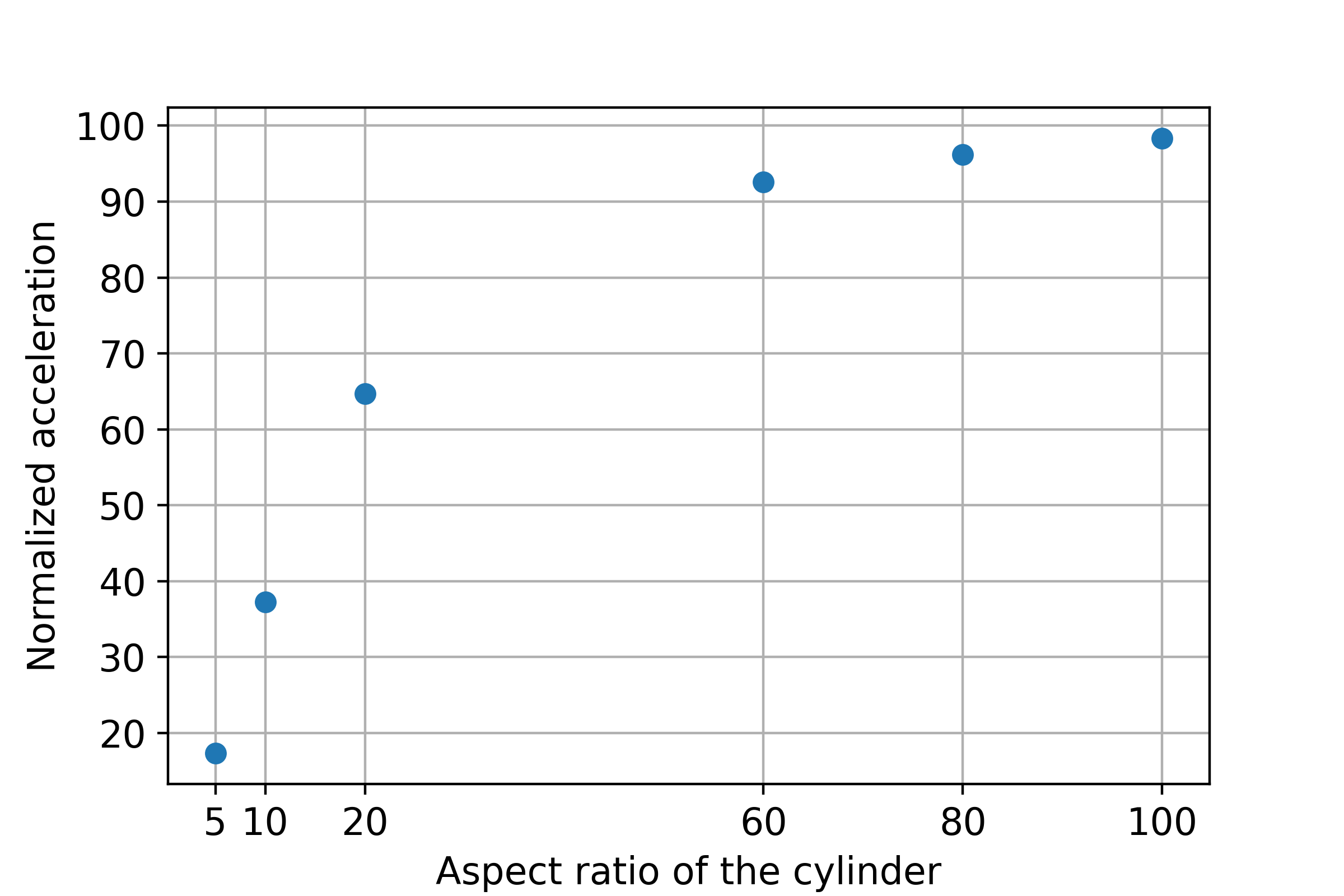}
    \caption{Simulated initial acceleration (normalized by $g$) of  cylindrical air bubbles with different aspect ratios of $[5,\,10,\,20,\,60,\,80,\,100]$ in the water at the time $t=10^{-6}s$. All bubbles have a volume of $60mm^3$. The cylinder axes are aligned with the direction of gravity.}
    \label{fig:accPlot-Cylinders}
\end{figure}
This thought experiment also indicates that much higher accelerations than $2g$ must be possible. In fact, we performed simulations for artificial bubbles of cylindrical shapes, with the cylinder axes aligned with the direction of gravity. For those we observed initial accelerations up to $100g$ for very slender cylinders, see Fig.~\ref{fig:accPlot-Cylinders}. This is not surprising, as it fits very well to the intuitive concept of added mass: Because of the small projected area in the rise direction, such a slender bubble only has to displace a small amount of liquid. Indeed, it is known that the added mass coefficient for slender bodies can be very small; see, e.g., 
\cite{brennen1982review} and \cite{simcik2013added}. This has been recently investigated experimentally in \cite{mckee2019acceleration}. As one cannot use a bubble for this experiment, an approximate ellipsoid (spindle) made of extruded polystyrene foam ('styrofoam') has been used instead. This way, initial accelerations of about $6.3 g$ were obtained.

It would be interesting to extend the presented approach
to compute the initial acceleration of a bubble with changing volume. The latter may be due to phase change, appearing e.g.\ for cavitation bubbles or in boiling flows, mass transfer, possibly accompanied by chemical reactions as for CO$_2$-bubbles dissolving in alkaline solutions, or pressure jumps inducing compressibility effects.
The latter has been exploited in experiments by \cite{ohl2003added} to determine the change in added mass during such a pressure change. 
A theoretical study using potential flow theory for a compressible bubble can be found in \cite{de2016dynamics}. To obtain the initial acceleration for a compressible bubble with volume change in a similar fashion as we did in this paper requires significant modifications and will be left for future work.

\ack{
Dieter Bothe, Jun Liu, Tomislav Maric and Matthias Niethammer gratefully acknowledge financial support by the Deutsche Forschungsgemeinschaft (DFG, German Research Foundation) - Project-ID 265191195 - SFB 1194.}
The numerical simulations for this research were carried out on the Lichtenberg high-performance computer of the TU Darmstadt. 
%


\bibliographystyle{RS}
\bibliography{jfm}

\end{document}